\documentclass[11pt, letterpaper]{article}
\usepackage[left=1in, right=1in, top=1in, bottom=1in]{geometry}
\usepackage[utf8]{inputenc}
\usepackage{natbib}
\usepackage{graphicx}
\usepackage{hyperref}
\usepackage{subcaption}

\usepackage{bm}
\usepackage{dcolumn}
\usepackage{ascmac}
\usepackage{amsmath,amssymb}

\newcommand{\pd}[2]{{\frac{\partial {#1}}{\partial {#2}}}}

\newcommand{\ci}{{\rm{i}}}
\newcommand{\Rey}{{Re}}
\newcommand{\Real}{{\text{Re}}}
\newcommand{\Imag}{{\text{Im}}}

\hypersetup{
    colorlinks = true,
    urlcolor   = blue,
    citecolor  = black,
}

\begin{document}

\title{\textbf{Triangular instability of a strained Batchelor vortex}} 

\author{
  A. S. P. Ayapilla\thanks{Corresponding author: \href{mailto:ayapilla.asp@scrc.iir.isct.ac.jp}{ayapilla.asp@scrc.iir.isct.ac.jp}. 
  Current address: Supercomputing Research Center, Institute of Integrated Research, Institute of Science Tokyo, Tokyo 152-8550, Japan}\\
  {\small Graduate School of Information Sciences, Tohoku University, Sendai 980-8579, Japan}
  \and
  Y. Hattori\\
  {\small Institute of Fluid Science, Tohoku University, Sendai 980-8577, Japan}
  \and
  S. Le Diz\`es\\
  {\small Aix Marseille Universit\'e, CNRS, Centrale M\'editerran\'ee, IRPHE, Marseille, France}
}
\date{}

\maketitle

\normalsize

\vspace{-18pt} 

\begin{abstract}
We investigate the triangular instability of a Batchelor vortex subjected to a stationary triangular strain field generated by three satellite vortices, in the presence of weak axial flow. The analysis combines theoretical predictions with numerical simulations. Theoretically, the instability arises from resonant coupling between two quasi-neutral Kelvin modes with azimuthal wavenumbers $m$ and $m+3$ with the background strain. Numerically, we solve the linearized Navier–Stokes equations around a quasi-steady base flow to identify the most unstable modes, and compare their growth rates and frequencies with theoretical predictions for a Reynolds number $\Rey = 10^4$ and a straining strength $\epsilon = 0.008$. In the absence of axial flow, only the mode pair $(m_A, m_B) = (-1,2)$ (and its symmetric counterpart) is unstable. However, we show that additional combinations such as $(0,3)$, $(1,4)$, and $(2,5)$, which are otherwise strongly damped by the critical layer in the absence of axial flow, also become unstable once axial flow exceeds a certain threshold, as the critical-layer damping is significantly reduced. Furthermore, we show that the most unstable mode in the no-axial-flow case, originating from the second branch of $m = -1$ and the first branch of $m = 2$, becomes less unstable as axial flow increases. It is eventually overtaken by a mode from the first branches of both wavenumbers, which then remains the dominant unstable mode across a wide range of axial flow strengths, Reynolds numbers and straining strengths. A comprehensive instability diagram as a function of the axial flow parameter is presented.
\end{abstract}

\vspace{12pt} 

\section{Introduction}
Vortices are naturally prone to instabilities triggered by small disturbances, which can amplify through inviscid or viscous mechanisms either within the vortex core or via interactions with the surrounding flow. Understanding vortex stability is of great theoretical interest and practical importance across a wide range of natural and industrial flows, including geophysical and astrophysical systems, aircraft wakes, and various rotating flows. A detailed overview of vortex stability theory can be found in \citet{ash1995vortex}.

Among the various types of instabilities, short-wave instabilities occur at scales comparable to or smaller than the vortex core. These arise through parametric resonance, where two Kelvin modes of the underlying vortex become resonant with the vortex correction induced by the background strain. Depending on the nature of the background strain, two well-studied types of short-wave instabilities are: (1) the elliptic instability, where the strain induces a quadrupolar correction that causes elliptic deformation of the vortex streamlines; and (2) the curvature instability, where a dipolar correction is introduced.
The elliptic instability is a key mechanism by which vortex pairs, such as wingtip vortices of airplanes, develop instabilities, alongside the Crow instability \citep{crow1970stability, leweke_dynamics_2016}. In such cases, each vortex induces a quadrupolar correction on its counterpart, leading to elliptic bending of streamlines and the onset of elliptic instability. In contrast, curvature instability typically arises from the bending of vortex tubes, such as in curved filaments and vortex rings, where the curvature creates a dipolar correction.
\citet{moore1975instability} and \citet{tsai_stability_1976} were the first to explain this mechanism as a parametric resonance, for the Lamb-Oseen vortex and Rankine vortex, respectively. A thorough summary of the elliptic instability is provided by \citet{kerswell_elliptical_2002}, with notable contributions from \citet{pierrehumbert1986universal, bayly1986three, leweke1998cooperative, le2002theoretical, meunier2005elliptic, schaeffer2010nonlinear, blanco-rodriguez_elliptic_2016, lacaze2007elliptic} among others.
Curvature instability was first observed in a helical vortex filament by \citet{widnall_stability_1972}, along with long-wave and mutual-inductance instability modes. Further theoretical and numerical studies on vortex rings and helical vortices include works by \citet{hattori_short-wavelength_2003, fukumoto_curvature_2005, blanco-rodriguez_curvature_2017, hattori_numerical_2019, xu2025helical} among others.

Multipolar short-wave instabilities, involving strain of order higher than two or non-axisymmetric vortices with higher symmetries, have received comparatively little attention. Long-lived non-axisymmetric vortices—such as dipoles, tripoles, and higher-order multipoles—have been observed in rotating turbulent flows through experiments and simulations \citep{hopfinger1993vortices, carnevale1994emergence, rossi1997quasi, dritschel1998persistence}. There is also evidence of Kelvin waves emerging on vortex filaments in transitional flows \citep{arendt1998kelvin}, with elliptic and multipolar instabilities potentially explaining their appearance and providing a mechanism for rich vortex dynamics of turbulent flows and transition. Multipolar instability has been studied theoretically and numerically for Rankine vortices \citep{le1999short, le2000three, eloy_stability_2001}, with experimental evidence of triangular instability reported in rotating cylinder configurations \citep{eloy2000experimental, eloy_elliptic_2003}. However, Rankine vortices are idealized, whereas models with continuous vorticity distributions, such as Lamb–Oseen and Batchelor vortices—self-similar solutions of the Navier–Stokes equations—offer a more realistic representation of vortices in wakes and rotating flows, with multipolar instability particularly having potential implications in rotor wakes such as wind turbines and ship propellers. This highlights the importance of studying multipolar instabilities in these models. Additionally, continuous vorticity profiles give rise to critical layers, which suppress many Kelvin modes and significantly affect instability behaviour \citep{Sipp03, fabre_viscous_2004, LeDizes04, dizes_asymptotic_2005, fabre_kelvin_2006}.

In a recent study \citep{triangjfm2025}, we presented the first evidence of triangular instability in a Lamb–Oseen vortex and characterized its behaviour both theoretically and numerically. For brevity, we refer to \cite{triangjfm2025} as AHL25. However, it is important to extend this analysis to the Batchelor vortex. In reality, vortices in wakes often possess a non-zero axial flow component, and the Batchelor vortex is recognized as a better model for trailing vortices in aircraft wakes \citep{batchelor1964axial}. Rotor wakes such as in Joukowski’s rotor wake model, typically consist of a hub vortex, root vortices, and tip vortices, where the hub vortex generally possesses axial flow. Therefore, investigating the effect of axial flow on the triangular instability is crucial. The influence of axial flow on the elliptic instability has been examined for a Rankine vortex by \citet{lacaze2005elliptic} and for the Batchelor vortex by \citet{lacaze2007elliptic}. These works showed that adding a small axial flow slightly modifies the dispersion curves and breaks the symmetry between Kelvin modes, leading to new resonant modes and altering how the critical layer affects the modes. For example, beyond the well-known stationary helical modes $m = 1$ and $-1$ in the elliptic instability of the Lamb–Oseen vortex, modes with $m = 0$, $2$, and other pairs were also shown to become unstable. The asymptotic behaviour of the Kelvin modes of the Batchelor vortex with small axial flow, in the large axial wavenumber limit, was analyzed using a Wentzel–Kramers–Brillouin (WKB) approach by \cite{dizes_asymptotic_2005}. The effects of axial flow on curvature instability have been studied in various configurations: for a helical vortex tube by \citet{hattori2012effects, hattori2014modal}, for a curved Batchelor vortex by \citet{blanco-rodriguez_curvature_2017}, and in a vortex ring with swirl by \citet{hattori_numerical_2019}, using numerical stability analysis.

The unstrained Batchelor vortex (or $q$-vortex) can become unstable when the axial flow is sufficiently strong (see \citet{ash1995vortex} for a review). Inviscid unstable modes were first computed numerically by \citet{lessen1974stabilitya} for swirl parameters $q < 1.5$, which corresponds to an axial flow parameter $W_0 > 0.66$. \citet{leibovich1983sufficient} later derived a sufficient condition for inviscid instability given by $q < \sqrt{2}$. Extending this further, \citet{heaton2007centre} demonstrated that weakly unstable inviscid modes can exist up to $q \approx 2.31$, i.e., for $W_0 > 0.43$. The Batchelor vortex also supports unstable viscous modes. These were first identified by \citet{khorrami1991viscous}, and later, a distinct class known as viscous centre modes localized near the vortex axis was obtained by \citet{fabre_viscous_2004} for low swirl values. An asymptotic description of these modes was provided by \citet{le2007large}, where the authors showed that the Batchelor vortex becomes unstable for any positive axial flow if the Reynolds number is sufficiently high. The full stability curve was further analyzed by \citet{fabre2008viscous}, who derived an expression for the critical swirl number as a function of Reynolds number, below which viscous centre modes become unstable. These asymptotic results will later be used in this work to analyze the competition between the viscous instability of the unstrained Batchelor vortex and the triangular instability of the strained vortex.

Capitalizing on this motivation, our objective in this study is to investigate how axial flow influences the triangular instability of a strained Batchelor vortex. In AHL25, we demonstrated that the triangular instability in the Lamb–Oseen vortex was restricted to the azimuthal wavenumber pair $m = 1$ and $-2$ (or $m = -1$ and $2$). Here, we expect this picture to change with the inclusion of small axial flow, which is the focus of the present investigation. We restrict attention to small axial flow values to ensure that triangular instability modes, if present, can become more unstable than the inviscid or the viscous centre modes previously discussed. As in AHL25, we adopt a combined approach of theoretical linear stability analysis and numerical stability analysis based on linearized direct numerical simulations, hereafter referred to simply as DNS. In the DNS framework, we first compute the base flow and then solve the linearized Navier–Stokes equations about that base flow to identify the most unstable mode. The theoretical framework involves deriving an approximate base flow under the assumption of a weak triangular straining field and using it to formulate the linear perturbation equations for the strained Batchelor vortex. These equations are then used to calculate the growth rates of the resonant modes. To identify these resonant modes, we solve the eigenvalue problem for the unperturbed vortex.
 
The paper is structured as follows. Section~\ref{sec:numerical_methodology} describes the numerical framework, beginning with the flow set-up in section~\ref{subsec:flow_setup}, followed by the governing equations for the base flow and linear stability analysis, and the numerical procedure in section~\ref{subsec:goveqn}. We then present the details of the numerical simulations and theoretical formulation of the base flow in section~\ref{sec:baseflow}, introduce the perturbation equations, and define the resonance condition for triangular instability in section~\ref{sec:perteq}. 
Section~\ref{sec:kelvinmodes} examines the Kelvin modes of the Batchelor vortex. In section~\ref{subsec:kelvinmodes_num}, we outline the numerical approach for computing the Kelvin modes of the unstrained vortex. In section~\ref{subsec:asymptotic}, we classify Kelvin modes using the large-$k$ asymptotic framework of \citet{dizes_asymptotic_2005} and identify the resonance domain in which triangular instability can arise.
Section~\ref{sec:tri_instab_char} analyses the characteristics of triangular instability. We first derive the theoretical growth-rate expression for resonant modes in section~\ref{subsec:theorymath}. In section~\ref{subsec:numresults}, we compare and validate the numerically obtained unstable modes, including their growth rates, against the theoretical predictions for selected axial flow values. In section~\ref{subsec:theoryresults}, we use the theoretical framework to conduct a parametric study exploring how the instability characteristics vary with axial flow, Reynolds number, and straining strength. 
Finally, section~\ref{sec:summary} summarizes the main findings and discusses potential directions for future work.

\section{Numerical Methodology}
\label{sec:numerical_methodology}
Let us recall that the numerical analysis consists of two main steps. First, we compute the quasi-steady base flow by solving the incompressible Navier–Stokes equations. Then for the linear stability analysis, we integrate the linearized Navier–Stokes equations around this base flow over a sufficiently long time to identify the most unstable mode, if one exists, for a specified axial wavenumber.

The framework for carrying out numerical simulations, including flow set-up, numerical methods, initial conditions, and simulation parameters are almost identical to those used in AHL25, with the key difference being that the current set-up includes axial flow. Below, we delve into the specific details and after formulating the flow set-up, we present the governing equations discretised and the numerical procedure used for the base flow and linear stability analysis. 

\subsection{Flow set-up}
\label{subsec:flow_setup}

\begin{figure}
  \centering
  \includegraphics[width=0.65\textwidth]{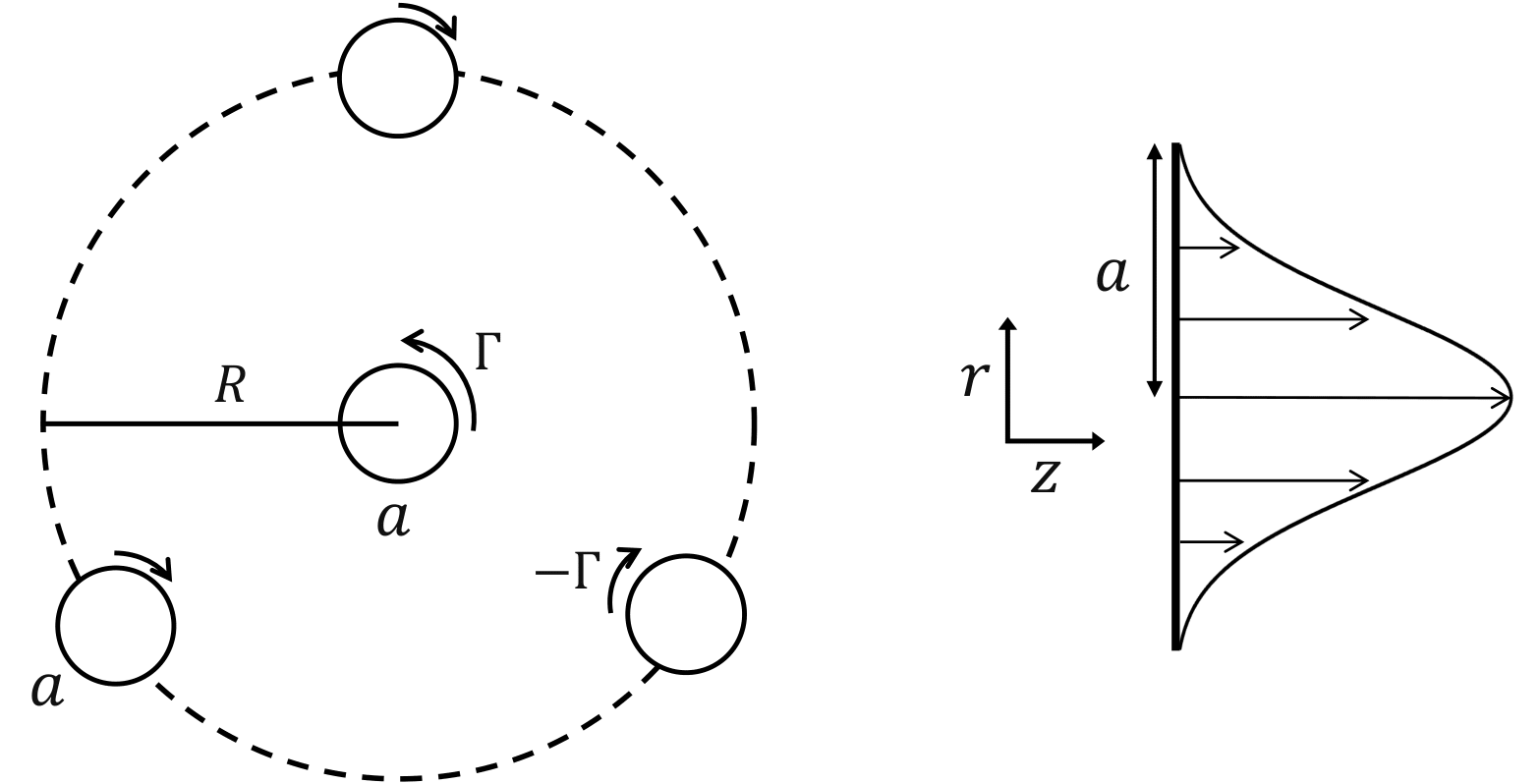}
  \caption{Flow configuration of the hub vortex and the three satellite vortices}
\label{fig:geometry}
\end{figure}

We consider the problem in a cylindrical coordinate system $(r,\theta,z)$. The initial configuration consists of a central hub Lamb--Oseen vortex located at $r=0$, symmetrically surrounded by three satellite Lamb--Oseen vortices in an incompressible, viscous fluid, as shown in figure~\ref{fig:geometry} in the $r$--$\theta$ plane. The three satellite vortices are located at a distance $R$ from the hub, separated by an angle of $2\pi/3$. Since the hub vortex is now assumed to have an axial flow in the $z$-direction with a Gaussian profile, as illustrated in figure \ref{fig:geometry}, it is modeled as a Batchelor vortex. The satellite vortices are modeled as Lamb-Oseen vortices and impose a triangular straining field on the hub vortex.  
The two-dimensional (2-D) vorticity distribution $\zeta_0(r)$ for all the vortices is therefore taken as 
\begin{align}
\zeta_0(r) = \dfrac{\Gamma_i}{\pi a_i^2} \exp\left(-\dfrac{|r - r_c|^2}{a_i^2} \right),
\label{eq2_1}
\end{align}
where $a_i$ is the initial radius of a vortex, $\Gamma_i$ is the initial circulation, and $r_c$ is the radial location of the vortex centre.

The axial velocity component $\overline{W}_0(r)$ of the Batchelor hub vortex is taken as
\begin{align}
\overline{W}_0(r) = W_{0,i} \exp\left(-\dfrac{r^2}{a_i^2} \right),
\label{eq2_3}
\end{align}
where $W_{0,i}$ is a free parameter that represents the strength of the axial flow. The subscript $i$ indicates that it corresponds to the initial state of the system. It is related to the swirl number $S$—defined as the ratio of the maximum azimuthal to axial velocity—by $W_{0,i} = 0.638/S$ \citep{lessen1974stabilitya}. 

The Reynolds number of the flow $\Rey$, is defined as
\begin{equation}
\Rey = \frac{\Gamma_i}{2 \pi \nu},
\end{equation}
where $\nu$ is the kinematic viscosity.

\subsection{Governing equations, domain and discretisation}
\label{subsec:goveqn}

The governing equations for both the base flow and linear stability analysis are derived from the three-dimensional (3-D) incompressible Navier--Stokes equations in cylindrical coordinates \((r, \theta, z)\) for the velocity field \(\mathbf{U} = (U, V, W)\) and pressure \(P\) given by: 
\begin{align}
&\frac{\partial U}{\partial t}
= -U \frac{\partial U}{\partial r}
  - \frac{V}{r} \frac{\partial U}{\partial \theta}
  - W \frac{\partial U}{\partial z}
  + \frac{V^2}{r}
  - \frac{\partial P}{\partial r}
  + \nu \left[ \left( \nabla^2 - \frac{1}{r^2} \right) U
  - \frac{2}{r^2} \frac{\partial V}{\partial \theta} \right],
\label{eqgov_1}\\[6pt]
&\frac{\partial V}{\partial t}
= -U \frac{\partial V}{\partial r}
  - \frac{V}{r} \frac{\partial V}{\partial \theta}
  - W \frac{\partial V}{\partial z}
  - \frac{UV}{r}
  - \frac{1}{r} \frac{\partial P}{\partial \theta}
  + \nu \left[ \left( \nabla^2 - \frac{1}{r^2} \right) V
  + \frac{2}{r^2} \frac{\partial U}{\partial \theta} \right],
\label{eqgov_2}\\[6pt]
&\frac{\partial W}{\partial t}
= -U \frac{\partial W}{\partial r}
  - \frac{V}{r} \frac{\partial W}{\partial \theta}
  - W \frac{\partial W}{\partial z}
  - \frac{\partial P}{\partial z}
  + \nu \nabla^2 W,
\label{eqgov_3}\\[6pt]
&\frac{\partial U}{\partial r}
  + \frac{U}{r}
  + \frac{1}{r} \frac{\partial V}{\partial \theta}
  + \frac{\partial W}{\partial z}
  = 0.
\label{eqgov_4}
\end{align}

\subsubsection{Base flow}
\label{subsubsec:baseflow}

It is important to note that although the base flow includes an axial velocity component, none of the three velocity components exhibit any variation along the axial ($z$) direction. The axial flow is therefore decoupled from the dynamics of the other two velocity components, and the evolution of the flow in the $r$–$\theta$ plane and $z$-direction are independent of each other. Accordingly, the governing equations are obtained by setting all partial derivatives with respect to $z$ in equations \eqref{eqgov_1}–\eqref{eqgov_4} to zero, and the resulting system is solved numerically.

The computational grid for the base flow is constructed in two dimensions, with discretisation performed only in the radial and azimuthal directions. A sixth-order compact finite difference scheme \citep{lele1992compact} is used in $r$, and a Fourier spectral method is employed in $\theta$, with periodic boundary conditions. The numerical domain spans $-1000 \leq r \leq 1000$, with the grid extended to negative $r$ to avoid the singularity at $r = 0$. As will be shown in section~\ref{sec:baseflow}, the hub--satellite vortex system is confined to $r < 10$; nevertheless, a large radial extent of 1000 is adopted to ensure numerical convergence. To keep the computation efficient, a non-uniform radial grid is adopted, using 695 points with a fine grid spacing of 0.0475 near the vortices and a coarser spacing farther away. In the azimuthal direction, 512 Fourier modes are used. The Poisson equation for pressure is decomposed into a set of ordinary differential equations for each Fourier mode and solved using the same sixth-order compact scheme. For temporal discretisation, the Crank–Nicolson scheme is applied to the viscous terms, while the second-order Adams–Bashforth method is used for the remaining terms. Further details on numerical methods and how the accuracy of the numerical simulations depends on the grid resolution are provided in Appendices A and B, respectively, of \citet{hattori_numerical_2019}.

\subsubsection{Linear stability analysis}
\label{subsubsec:linearstabilityanalysis}

The linearized DNS, on the other hand, is extended to three dimensions $(r, \theta, z)$, where a single normal mode with a specified axial wavenumber is considered in the $z$ direction. Specifically, if the base flow is $[U(r,\theta), V(r,\theta), W(r,\theta)]$, then the linearized Navier–Stokes equations about the base flow for disturbances $u'$, $v'$, $w'$, and $p'$  which are functions of $r, \theta$, and $z$  are given by
\begin{eqnarray}
\pd{u'}{t} &=& -U\pd{u'}{r} -\frac{V}{r}\pd{u'}{\theta} - W\pd{u'}{z} + \frac{2Vv'}{r}
- u'\pd{U}{r} -\frac{v'}{r}\pd{U}{\theta} \nonumber \\
& & -\pd{p'}{r} 
+ \frac{1}{Re} \left[\left(\nabla^2 -\frac{1}{r^2}\right) u' - \frac{2}{r^2} \pd{v'}{\theta}\right], \label{eqlin1}\\
\pd{v'}{t} &=& -U\pd{v'}{r} -\frac{V}{r}\pd{v'}{\theta} - W\pd{v'}{z}
- u'\pd{V}{r} -\frac{v'}{r}\pd{V}{\theta} - \frac{Uv' + u'V}{r} \nonumber \\
& & -\frac{1}{r}\pd{p'}{\theta} 
+ \frac{1}{Re} \left[\left(\nabla^2 -\frac{1}{r^2}\right) v' + \frac{2}{r^2} \pd{u'}{\theta}\right], \label{eqlin2}\\
\pd{w'}{t} &=& -U\pd{w'}{r} -\frac{V}{r}\pd{w'}{\theta} - W\pd{w'}{z}
- u'\pd{W}{r} - \frac{v'}{r} \pd{W}{\theta}
- \pd{p'}{z} 
+ \frac{1}{Re} \nabla^2 w'. \label{eqlin3}
\end{eqnarray}
where $\nabla^2 = \partial_r^2 + (1/r)\partial_r + (1/r^2) \partial^2_{\theta} + \partial^2_{z}$.  
These equations are discretised and solved numerically on the same $(r-\theta)$ grid as the base flow. The single wave in the $z$-direction is represented by assuming periodic boundary conditions as
\begin{eqnarray*}
F(t, r, \theta, z) = e^{\ci k z} \sum_{m} \hat{F}_{m} (t,r) e^{\ci m\theta}  
\end{eqnarray*}
for the disturbance fields $u$, $v$, $w$, and $p$, and thus rendering the equations separable in $z$. The same temporal discretisation as the base flow is used.

\section{Base flow}
\label{sec:baseflow}
\subsection{Numerical simulations}
\label{subsec:baseflow_num}
This section outlines the initial conditions and flow parameters used in the numerical simulations to obtain the quasi-steady state of the system, which then serves as the base flow for the subsequent linearized simulations. We restrict our analysis to a non-rotating reference frame, and to achieve this, we assign the initial circulation of the hub vortex to be equal and have opposite sign to those of satellite vortices. So for the hub vortex, we take $\Gamma_i = \Gamma = 2\pi$. The circulation of each satellite vortex is thus set to $-2\pi$. This choice ensures that the net induced velocity at the centre of any vortex due to the others vanishes (negligible, although not perfectly zero), allowing the vortex system to remain stationary. We set $\Rey = 1000$. Note that the quasi-steady state is anyway not expected to depend on the Reynolds number \citep{LeDizes02a} and we chose a lower value of Reynolds number to ensure that the quasi-steady state is reached more rapidly. Under these conditions, the system consisting of the hub and satellite vortices evolves toward a quasi-steady state once all vortices reach equilibrium under their mutual interaction. We reiterate that the relaxation mechanism is identical to that of the Lamb–Oseen case, since the dynamics in the 2-D plane are independent of the axial flow. The axial velocity field also evolves independently according to the same governing equation as the axial vorticity, with the axial-flow strength $W_{0,i}$ as the only varying parameter. Consequently, the axial flow reaches its quasi-steady state in the same manner as the planar flow field. For verification of the quasi-steady state, we refer the reader to section 3 (figures 4 and 5) of AHL25 and do not repeat it here.

The initial core radii of both the hub and satellite vortices, $a_i$, gradually increase during the relaxation process, stabilizing at a larger value $a$ given as
\begin{equation}
  a^2(t) = a_i^2 + 4\nu t. \label{eq_numb1}
\end{equation}
We begin the simulations with a core radius of $a_i = 0.8$ for both hub and satellite vortices, and terminated it when the core radius of the hub vortex reaches $a = 1$, which occurs after the quasi-steady state is established. The value of $a = 1$ is used to non-dimensionalise all spatial quantities in the theoretical analysis, presented in the next section. The distance between the hub and satellite vortices is set to $R = 5$, and it remains nearly constant throughout the relaxation process. 
Therefore, at the quasi-steady state, we have $a = 1$, $R = 5$, and at the leading order, the axial vorticity and the axial velocity of the hub vortex take the form
\begin{align}
  \zeta_0 &= 2e^{-r^2}, \label{eq_numb2} \\
  \overline{W}_0 &= W_0 e^{-r^2}, \label{eq_numb3} 
\end{align}
The initial axial flow strength $W_{0,i}$ is set so that a desired target $W_0$ is achieved at the quasi-steady state. 

\subsection{Theoretical description}
\label{subsec:baseflow_theo}
This section outlines the theoretical formulation of the base flow. While the numerical simulations account for the full configuration—including the hub vortex and the three satellite vortices—the theoretical analysis focuses solely on the hub vortex. The goal is to derive an approximate expression for the hub vortex’s velocity field influenced by the surrounding satellites. We omit the full derivation and refer the reader to section 3 of AHL25 and earlier works by \citet{moffatt_stretched_1994}, \citet{jimenez_structure_1996}, and \citet{dizes_non-axisymmetric_2000}. 

An asymptotic approximation of the base flow velocity field is obtained in the limit of small $a/R$ by modeling the satellite vortices as line vortices, introducing the small parameter $\epsilon = (a/R)^3$. For the quasi-steady values of $a$ and $R$ in our set-up, this gives $\epsilon = 0.008$. In practical applications such as wind turbines and marine propellers, the ratio of the hub radius to the blade length, $a/R$, typically lies in the range $0.05$–$0.2$. Moreover, the effective hub-vortex radius may be even larger in practice, as it forms through the merging of vortices generated near the blade roots. In contrast, tip vortices are substantially smaller—often of the order of $1/100$ of the blade length, or even less—which further justifies the use of a line-vortex approximation for modeling the tip vortices. Overall, it can be reasonably argued that the value $\epsilon = 0.008$ lies within a realistic range for practical rotors and is sufficiently large to serve as a useful starting point for investigating triangular instability from a theoretical perspective.

\begin{figure}
  \centering
  \includegraphics[width=0.7\textwidth]{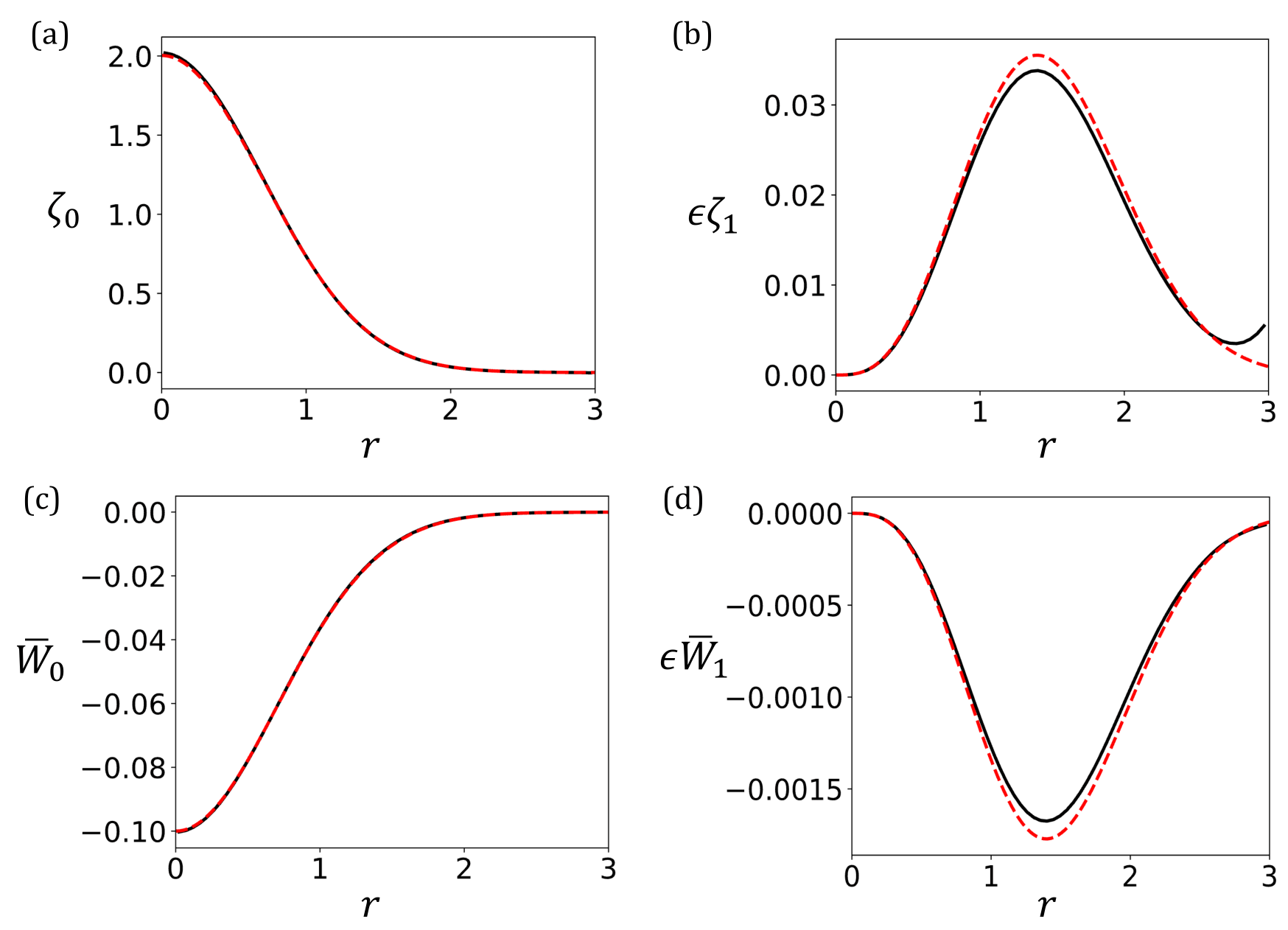}
  \caption{Comparison of the axial vorticity and axial velocity of the quasi-steady base flow obtained from DNS (solid black line) and theory (red dashed line) for $\Rey = 1000$ and $\epsilon = 0.008$. Panels (a) and (b) show the leading-order term and the triangular-straining correction of the axial vorticity, respectively. Panels (c) and (d) present the corresponding leading-order and correction terms for the axial velocity.}
\label{fig:bf_comp}
\end{figure}

The velocity field $(\overline{U}, \overline{V}, \overline{W})$ and the vorticity field $\zeta$ for the triangularly strained Batchelor vortex can be approximated as
\begin{align}
  \overline{U}(r,\theta) &= \epsilon \frac{3f(r)}{r} \cos{3\theta} + O(\epsilon^2), \label{eq1} \\
  \overline{V}(r,\theta) &= r\Omega_0(r) - \epsilon f'(r)\sin{3\theta} + O(\epsilon^2), \label{eq2} \\
  \overline{W}(r,\theta) &= \overline{W}_0 + \epsilon \overline{W}_1 + O(\epsilon^2), \label{eq3} \\
  \zeta(r,\theta) &= \zeta_0 + \epsilon \zeta_1 + O(\epsilon^2), \label{eq4}
\end{align}
where
\begin{align}
  \overline{W}_0 &= W_0 e^{-r^2}, \label{eq5} \\
  \overline{W}_1 &= \frac{W_0}{2}g(r) \sin{3\theta}, \label{eq7} \\
  \zeta_0 &= 2e^{-r^2}, \label{eq6} \\
  \zeta_1 &= g(r) \sin{3\theta}, \label{eq8}
\end{align}
and
\begin{equation}
  g(r) = \frac{4r^2 f(r)}{1 - e^{r^2}}. \label{eq10}
\end{equation}
Here, $\Omega_0(r)$ denotes the angular velocity, given by $\Omega_0(r) = (1 - e^{-r^2}) / r^2$, and $f(r)$ satisfies the second-order linear ordinary differential equation
\begin{equation}
   {f}'' + r^{-1}{f}' - \left(9r^{-2} + \frac{4r^2}{\left(1 - e^{r^2} \right)}\right)f = 0,
\label{eq17a}
\end{equation}
with the boundary conditions,
\begin{equation}
    f(r) \sim  s_0r^3, \; \; \; at \; \; \; r \rightarrow 0, 
\label{eq18}
\end{equation}
\begin{equation}
  f(r) \sim  r^3, \; \; \; at \; \; \; r \rightarrow \infty.
\label{eq19}
\end{equation}

The constant $s_0\approx  1.7724$ appearing in these equations is a numerical constant derived from the integration of equation (\ref{eq17a}).  It corresponds to the amplitude of the straining field in the vortex centre, normalized by the amplitude of the straining field at the same point in the absence of the vortex. 

Next, we briefly compare and validate the numerically obtained base flow against the theoretical prediction. Figure~\ref{fig:bf_comp} shows a comparison of the axial vorticity and axial velocity at the quasi-steady state from DNS and theory. The agreement is excellent, further confirming that the 2-D flow and the axial flow evolve independently while obeying the same governing equations. For this example, $W_{0,i}$ was chosen such that the quasi-steady value becomes $W_0 = -0.1$, although, as noted earlier, $W_{0,i}$ may be freely selected to achieve any desired $W_0$ at the quasi-steady state. Panels (a) and (b) of figure~\ref{fig:bf_comp} correspond to the same results presented in figure~5 of AHL25. Finally, we note that in figure~\ref{fig:bf_comp}(b), the increase in the numerically obtained correction term of the vorticity near $r \approx 3$ arises from the presence of the satellite vortices. No such feature appears in panel~(d), since no axial flow is imposed on the satellite vortices.

\section{Linear perturbation equations and triangular instability mechanism}
\label{sec:perteq}
Using equations~(\ref{eq1})-(\ref{eq3}) for the base flow, the linearized Navier-Stokes equations for the velocity--pressure field   $\mathbf{u} = (u,v,w,p)$ of the perturbations can be written in the form 

\begin{equation}
  \left( \mathcal{L}\frac{\partial}{\partial t} + \mathcal{P}\frac{\partial}{\partial z} + \mathcal{M} - \frac{\mathcal{V}}{\Rey} \right)\mathbf{u} = \epsilon\left( e^{3\mathrm{i}\theta}\mathcal{N} + e^{-3\mathrm{i}\theta}\overline{\mathcal{N}} \right)\mathbf{u},
\label{eq20}
\end{equation}
where the matrices $\mathcal{L}$, $\mathcal{P}$, $\mathcal{M}$, $\mathcal{V}$, $\mathcal{N}$  are given in appendix \ref{appA}.

\label{mechanism}
The mechanism for the growth of perturbations is due to the resonance of two quasi-neutral Kelvin modes of the Batchelor vortex with the triangular strain field. 
The Kelvin modes can be expressed as
\begin{equation}
\mathbf{u} = \mathbf{\Tilde{u}}(r)e^{\mathrm{i}(m\theta + kz - \omega t)},
\label{eq21}
\end{equation}
where $m$ is the azimuthal wavenumber, $k$ the axial wavenumber, $\omega$  the frequency, and $\mathbf{\Tilde{u}}(r)$  the eigenfunction field of the Kelvin mode. 
They satisfy the perturbation equations for the unstrained vortex,
{\begin{equation}
   \left( \omega\mathcal{L} - k\mathcal{P} + \mathrm{i}\mathcal{M}(m)-\frac{\mathrm{i}}{\Rey}\mathcal{V}(m,k)\right)\mathbf{\Tilde{u}} = 0, \label{eq22} 
\end{equation}
where $\mathcal{M}(m)$ and $\mathcal{V}(m,k)$ are obtained by replacing $\partial/\partial\theta$ with $\mathrm{i}m$ and $\partial/\partial z$ with $\mathrm{i}k$, in the matrix operators $\mathcal{M}$ and $\mathcal{V}$, given in appendix \ref{appA}.}

To identify the possible coupling of two Kelvin modes with the strain field, one should identify two neutral Kelvin modes 
$(\omega_A, k_A, m_A)$ and $(\omega_B, k_B, m_B)$ that satisfy
\begin{equation}
 \omega_A =  \omega_B,\;\;\; k_A = k_B,\;\;\;m_B-m_A=3.
\label{eq23}
\end{equation}

In the presence of viscosity, these modes are always damped, i.e., $\Imag(\omega) < 0$. While some modes become neutral as the Reynolds number tends to infinity, others remain damped due to the presence of a critical layer \citep{Sipp03}. 

\section{Kelvin modes of the Batchelor vortex}
\label{sec:kelvinmodes}

\subsection{Numerical computation of the Kelvin modes}
\label{subsec:kelvinmodes_num}

The Kelvin modes of the unstrained Batchelor vortex are computed by numerically solving the set of equations~\eqref{eq22}, along with the appropriate boundary conditions, which together form an eigenvalue problem.  At $r = \infty$, all the eigenfunctions $u$, $v$, $w$, and $p$ must decay exponentially to zero for $k \neq 0$ (the case $k = 0$ is not considered). At the vortex centre, $r = 0$, a regular singularity is encountered, and the asymptotic forms of the eigenfunctions can be derived as 
\begin{equation}
    \Tilde{u} = O(r), \quad  \Tilde{v} = O(r), \quad \Tilde{w} = O(1), \quad  \Tilde{p} = O(1),   
\end{equation}
for $m = 0$, and 
\begin{equation}
    \Tilde{u} = O(r^{|m|-1}), \quad  \Tilde{v} = O(r^{|m|-1}), \quad  \Tilde{w} = O(r^{|m|}), \quad  \Tilde{p} = O(r^{|m|}),   
\end{equation}
for $|m| \neq 0$. The numerical approach to solve this eigenvalue problem closely follows the method used in AHL25, with the only difference being the inclusion of a small axial flow in the present study. A Chebyshev spectral collocation method is employed, following the procedure outlined by \citet{fabre_viscous_2004}. The original domain $0 < r < \infty$ is extended symmetrically to $-\infty < r < \infty$ and mapped onto a contour in the complex-$r$ plane. The problem is then discretised in the Chebyshev domain $(-1, 1)$ using $2(N+1)$ collocation points, where a resolution of $N=200$ is found to be sufficient.
In the inviscid case, we apply a complex mapping similar to that used by \citet{fabre_viscous_2004}, defined by
\begin{equation}
 r = \frac{H\xi}{1-\xi^2} + \mathrm{i}\frac{A\xi}{\sqrt{1-\xi^2}},
\label{eqmap}
\end{equation}
where $H$ controls the radial distribution of the collocation points and $A$ sets the tilt of the contour into the complex plane.
We also exploit the parity properties of the eigenfunctions to optimize the computation. For odd azimuthal wavenumbers $m$, $\tilde{w}$ and $\tilde{p}$ are expanded in odd polynomials, while $\tilde{u}$ and $\tilde{v}$ use even polynomials. For even $m$, the roles are reversed, with $\tilde{w}$ and $\tilde{p}$ using even polynomials and $\tilde{u}$ and $\tilde{v}$ using odd polynomials.

\begin{figure}
  \centering
  \includegraphics[width=1.0\textwidth]{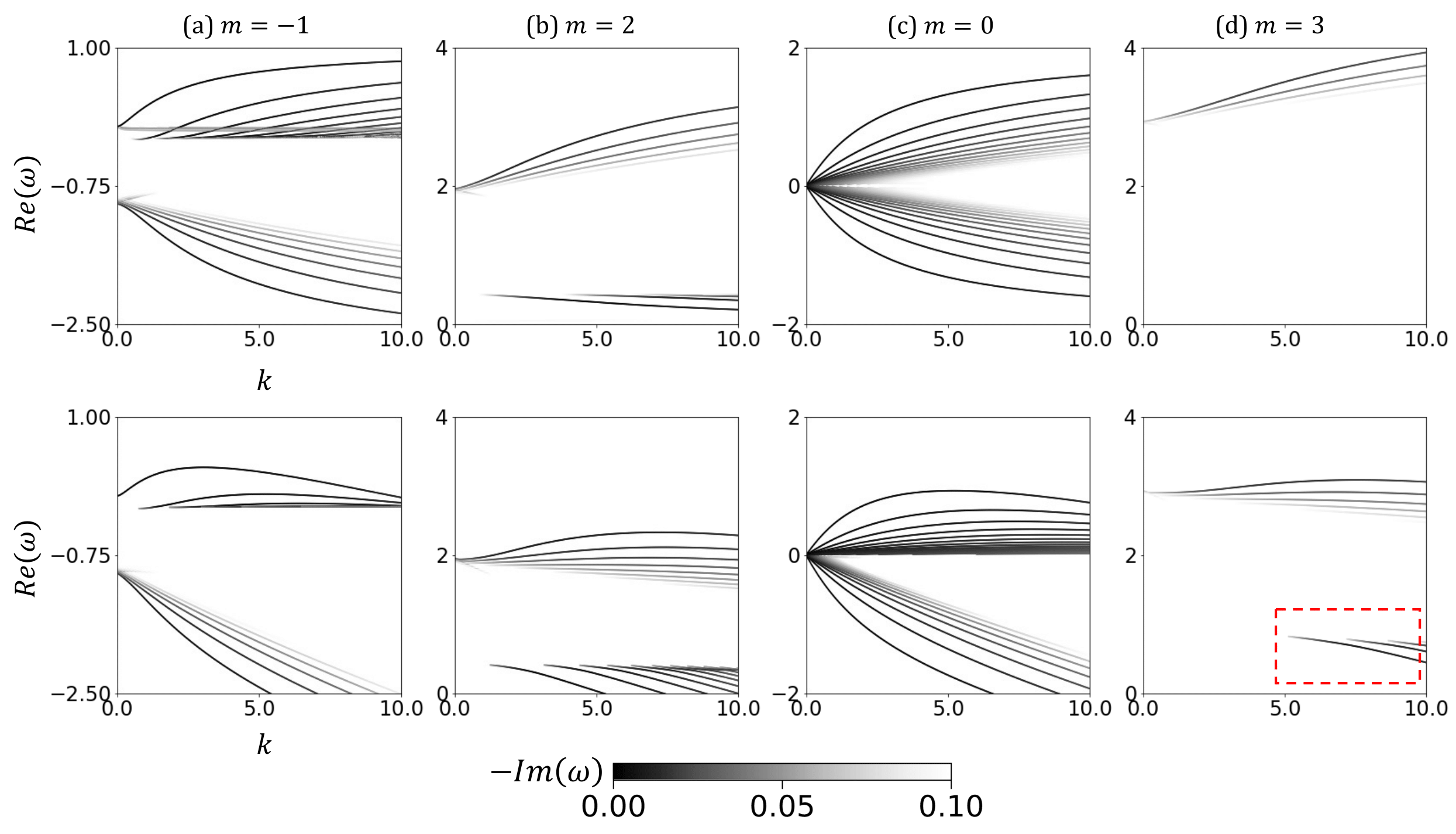}
  \caption{Effects of axial flow on the dispersion curves ($\Real(\omega)$ vs. $k$) of Kelvin modes of the Batchelor vortex, computed at $\Rey = 10^4$ by integration along the real axis. Each point represents a mode, with its greyscale intensity corresponding to the damping rate $-\Imag(\omega)$: darker points indicate lower damping ($=0$) and lighter points higher damping ($=0.1$). Top panels show results for $W_0 = 0$ (Lamb–Oseen vortex), and bottom panels for $W_0 = -0.1$. Results are shown for (a) $m = -1$, (b) $m = 2$, (c) $m = 0$, and (d) $m = 3$.}

\label{fig:batchelormodes_dispcurves}
\end{figure}

\begin{figure}
  \centering
  \includegraphics[width=1.0\textwidth]{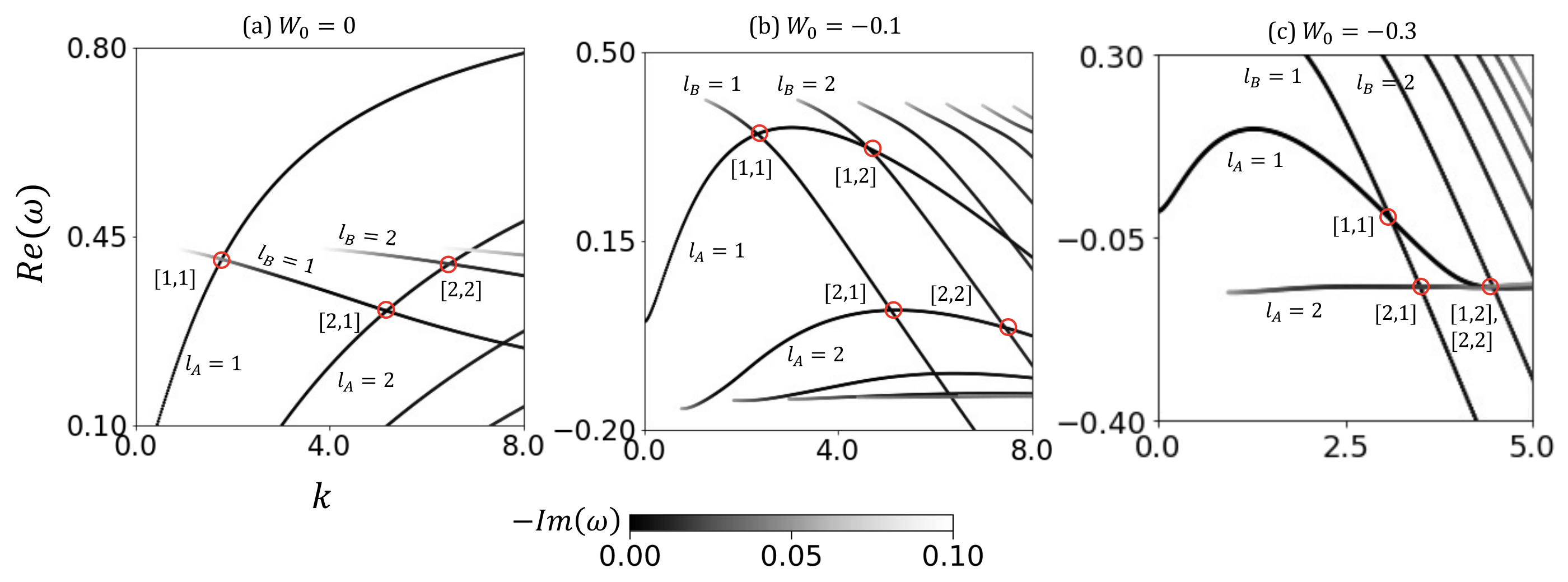}
  \caption{Effects of axial flow on the crossing points of the dispersion curves ($\Real(\omega)$ vs. $k$) for $m_A = -1$ and $m_B = 2$. Results are shown for (a) $W_0 = 0$, (b) $W_0 = -0.1$, and (c) $W_0 = -0.3$. Each point represents a mode, with its greyscale intensity corresponding to the damping rate $-\Imag(\omega)$: darker points indicate lower damping ($=0$) and lighter points higher damping ($=0.1$).}

\label{fig:batchelormodes_m-12comp}
\end{figure}

The effect of axial flow on the dispersion curves is illustrated in figure~\ref{fig:batchelormodes_dispcurves} for $m = -1$, $0$, $2$, and $3$. Throughout this paper, we focus on negative values of $W_0$, ensuring that the axial wavenumber $k$ remains positive and the signs of $m$ and $\Real(\omega)$ are preserved. As shown in the figure, increasing the magnitude of $W_0$ causes the dispersion curves to shift downward. For $m = 0$, the curves, which are symmetric about zero when $W_0 = 0$, lose their symmetry once axial flow is introduced. Another notable effect is observed for $m = 3$: at $W_0 = -0.1$, new branches (highlighted by the red-dashed rectangle) emerge that were previously suppressed by the critical layer when $W_0 = 0$. The appearance of these modes enables new resonance possibilities with $m = 0$ modes, potentially leading to triangular instability. Similarly, this happens for other azimuthal pairs such as $(1,4)$, $(2,5)$, and beyond. The nature of these modes will be discussed in the next section.

Figure~\ref{fig:batchelormodes_m-12comp} illustrates the effect of axial flow on the crossing points of the dispersion curves for $m = -1$ and $m = 2$. Before proceeding with the discussion, we first establish the notation that will be used throughout the paper to refer to modes constituting a resonant azimuthal wavenumber pair $(m_A, m_B)$. For example, the results shown in the figure correspond to the pair $(m_A, m_B) = (-1,2)$, where $m_A = -1$ and $m_B = 2$. The labels $[l_A, l_B]$ denote the respective branches of the dispersion curves to which the modes of $m_A$ and $m_B$ belong. For instance, $[l_A, l_B] = [1,2]$ indicates that the mode associated with $m_A$ lies on its first branch, while the mode for $m_B$ lies on its second branch. For brevity, we will sometimes use the notation $(m_A, m_B, [l_A, l_B])$ to refer to both the azimuthal wavenumbers of a resonant pair and their associated branches.

From figure~\ref{fig:batchelormodes_m-12comp}, it is evident that at $W_0 = 0$, some branches of $m = 2$ were heavily damped by the critical layer and could not form crossing points with $m = -1$. However, at $W_0 = -0.1$, these modes become less damped and are able to form crossing points. For instance, the resonant pair $[1,2]$ does not exist at $W_0 = 0$ but appears at $W_0 = -0.1$. Additionally, the critical-layer damping associated with the pairs $[1,1]$ and $[2,2]$ can be seen to weaken as $W_0$ decreases from $0$ to $-0.1$.

Another important observation is that the resonant modes originating from $m = -1$, which were undamped at $W_0 = 0$, can become damped at higher magnitudes of axial flow for specific values of $k$ and $W_0$. For example, at $W_0 = -0.3$, the modes corresponding to the pairs $[2,1]$ and $[2,2]$ are significantly damped by the critical layer. 

While these observations are qualitative, based on visual inspection of the dispersion curves, they clearly illustrate how axial flow alters the characteristics of resonant modes. A more detailed and quantitative classification of these effects will be provided in the next section.

\subsection{Large-$k$ asymptotic prediction of the Kelvin modes}
\label{subsec:asymptotic}

In this section, we shall discuss the asymptotic nature of the various kinds of modes that can exist for the Batchelor vortex in the inviscid limit. \citet{dizes_asymptotic_2005} developed an asymptotic framework based on WKB analysis to characterize the different types of Kelvin modes that arise in the limits of infinite Reynolds numbers and large axial wavenumbers. The essential characteristics of these modes can be determined by examining three functions: the epicyclic frequencies $\omega^{\pm}(r)$ and the critical frequency $\omega_c(r)$, defined as
\begin{equation}
\omega^{\pm}(r) = m\Omega_0(r) + k\overline{W}_0 \pm \sqrt{2\Omega_0(r)\zeta_0(r)},
\label{eqc1}
\end{equation}
\begin{equation}
\omega_c(r) = m\Omega_0(r) + k\overline{W}_0,
\label{eqc2}
\end{equation}
where $\overline{W}_0$ is given in equation~\eqref{eq5}, and $\Omega_0(r)$ and $\zeta_0(r)$ are defined in section~\ref{subsec:baseflow_theo}.

For given values of $m$, $k$, and $W_0$, these functions can be plotted to identify the frequency ranges corresponding to different mode types: (i) regular or singular neutral modes, and (ii) core or ring modes. In the case of the Lamb–Oseen vortex ($W_0 = 0$), the curves are independent of $k$, resulting in identical profiles for all axial wavenumbers. However, for the Batchelor vortex, each value of $k$ yields distinct $\omega^{\pm}(r)$ and $\omega_c(r)$ profiles, leading to a unique frequency range for each mode type.

\begin{figure}
  \centering
  \includegraphics[width=1.0\textwidth]{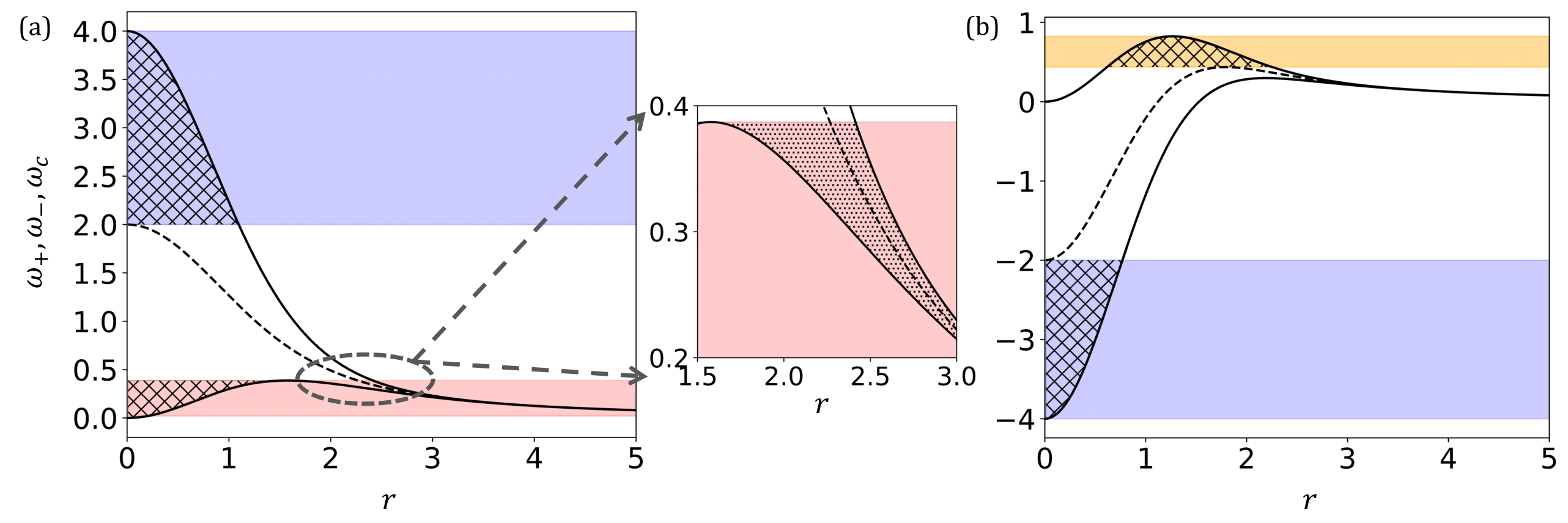}
  \caption{Plots of the epicyclic frequencies $\omega^+$ and $\omega^-$ (solid lines) and the critical frequency $\omega_c$ (dashed line) as functions of the radial coordinate $r$. Results are shown for (a) $m = 2$, $kW_0 = 0$, with an inset showing a zoomed-in view of the indicated region; (b) $m = 2$, $kW_0 = -4$. The blue and red shaded areas represent the domains of regular neutral core modes and singular neutral core modes, respectively. In (b), the yellow shaded region corresponds to regular ring modes. Hatched regions denote the regions where modes are localized. The dotted region in the inset of (a) highlights where singular neutral modes encounter a critical point given by $\omega_c$.}

\label{fig:batchelormodes_asymcurves}
\end{figure}

Figure~\ref{fig:batchelormodes_asymcurves} shows plots of $\omega^{\pm}(r)$ and $\omega_c(r)$ for $m = 2$ in two cases, $kW_0 = 0$ and $kW_0 = -4$, illustrating the domains where each type of mode can exist. To determine the nature of a mode with a given frequency, one can draw a horizontal line at that frequency value. The region enclosed between the curves $\omega^{\pm}(r)$ is where the modes are localized and have oscillatory behaviour, and decay exponentially to zero in the areas outside this region. The intersection points with the $\omega^{\pm}(r)$ curves are the WKB turning points. Intersections with the $\omega_c(r)$ curve mark critical point singularities in $r$. Based on the locations of these intersection points, the type of mode can be classified as follows.

\subsubsection*{Regular neutral modes vs. singular neutral modes}
\label{subsubsec:regvssing}

\begin{figure}
  \centering
  \includegraphics[width=1.0\textwidth]{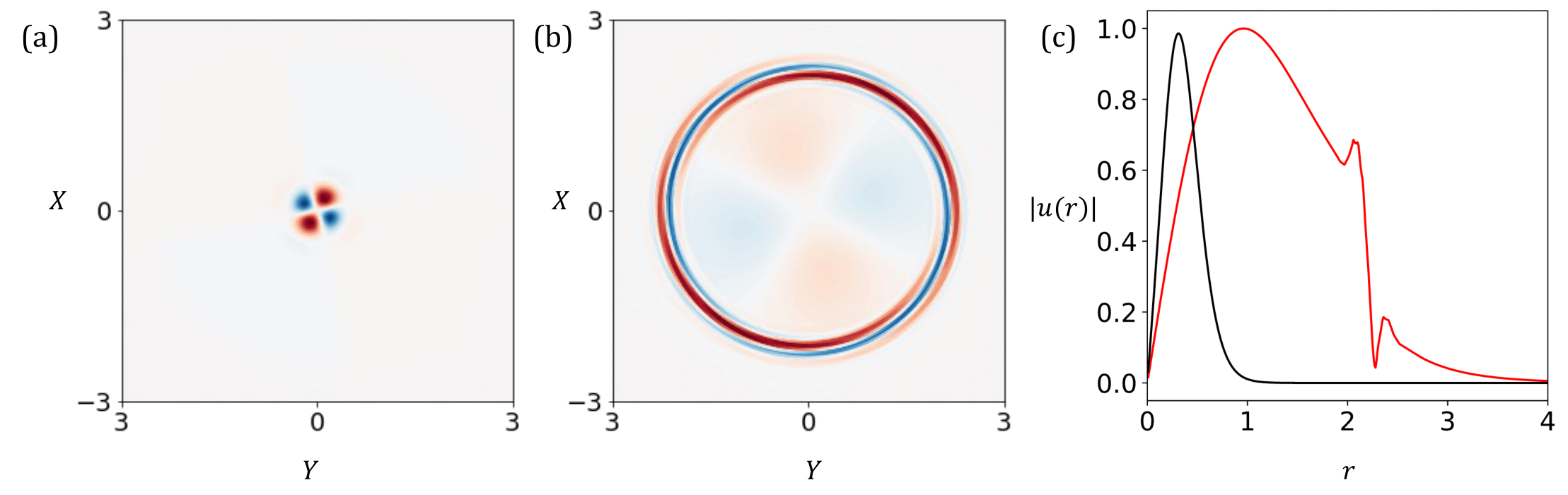}
  \caption{Comparison of the structures of a regular neutral mode and a singular neutral mode obtained at $Re = 10^4$, $m = 2$, $kW_0 = 0$. (a) Regular neutral mode: 2-D contours of axial disturbance vorticity for the mode with $k = 10.0$, $\Real(\omega) = 3.138$, $\Imag(\omega) = -0.020$; (b) singular neutral mode: $k = 1.76$, $\Real(\omega) = 0.407$, $\Imag(\omega) = -0.045$; (c) radial distribution of the absolute value of the eigenfunction $|u(r)|$, with the black curve corresponding to the mode in (a) and the red curve to the mode in (b).}
\label{fig:batchelormodes_regvscrit}
\end{figure}

A mode is classified as a regular neutral mode if it does not encounter any critical point singularities in $r$—that is, if a horizontal line drawn at its frequency does not intersect $\omega_c(r)$ anywhere. In figure~\ref{fig:batchelormodes_asymcurves}, regions where modes are regular are shaded in blue. 

However, if a mode encounters a critical point at large $r$—after passing through at least two turning points, as indicated by the red regions in the figure—it is classified as a singular neutral mode and undergoes asymptotically small critical-layer damping. Such modes are often referred to as critical-layer waves. When solving equation~\eqref{eq22}, the critical point can be avoided by deforming the integration contour into the complex plane, either above or below the critical point, depending on its trajectory in the complex plane \citep{LeDizes04}. Alternatively, the critical point can be regularized by introducing viscosity while remaining on the real axis. As a result, the structure of such modes is divided into two distinct radial regions: in the first region, the mode is oscillatory like a regular neutral mode, while in the second region (illustrated by the zoomed inset in figure~\ref{fig:batchelormodes_asymcurves}(a)), the mode exhibits additional oscillations induced by the regularization of the critical layer. The additional modes that appeared for $m = 3$ in the presence of axial flow, seen in figure~\ref{fig:batchelormodes_dispcurves}(d), are singular neutral modes. Singular neutral modes exist for all $m \neq 0$ in specific frequency ranges.

It is also insightful to compare the structures of a regular neutral mode with those of a singular neutral mode. Figure~\ref{fig:batchelormodes_regvscrit} illustrates this by showing the axial vorticity fields of one representative regular neutral mode from the blue region and one singular neutral mode from the red region in figure~\ref{fig:batchelormodes_asymcurves}(a). In the latter, additional oscillations caused by the critical layer regularization can be observed in figure~\ref{fig:batchelormodes_regvscrit}(c). The 2-D structure of the singular neutral mode, shown in figure~\ref{fig:batchelormodes_regvscrit}(b), is also characterized by spiral arms in the critical layer, while retaining a regular wave structure in the inner region similar to that of the regular neutral mode displayed in figure~\ref{fig:batchelormodes_regvscrit}(a). 

If a critical point is encountered before passing through at least two turning points, the mode is completely suppressed by the critical layer in the inviscid limit, although it can still exist as a strongly damped viscous mode at lower Reynolds numbers \citep{fabre_kelvin_2006}. The unshaded regions in figure~\ref{fig:batchelormodes_asymcurves} correspond to the domains where such modes exist. 

\subsubsection*{Core modes versus ring modes}
\label{subsubsec:corevsring}

\begin{figure}
  \centering
  \includegraphics[width=1.0\textwidth]{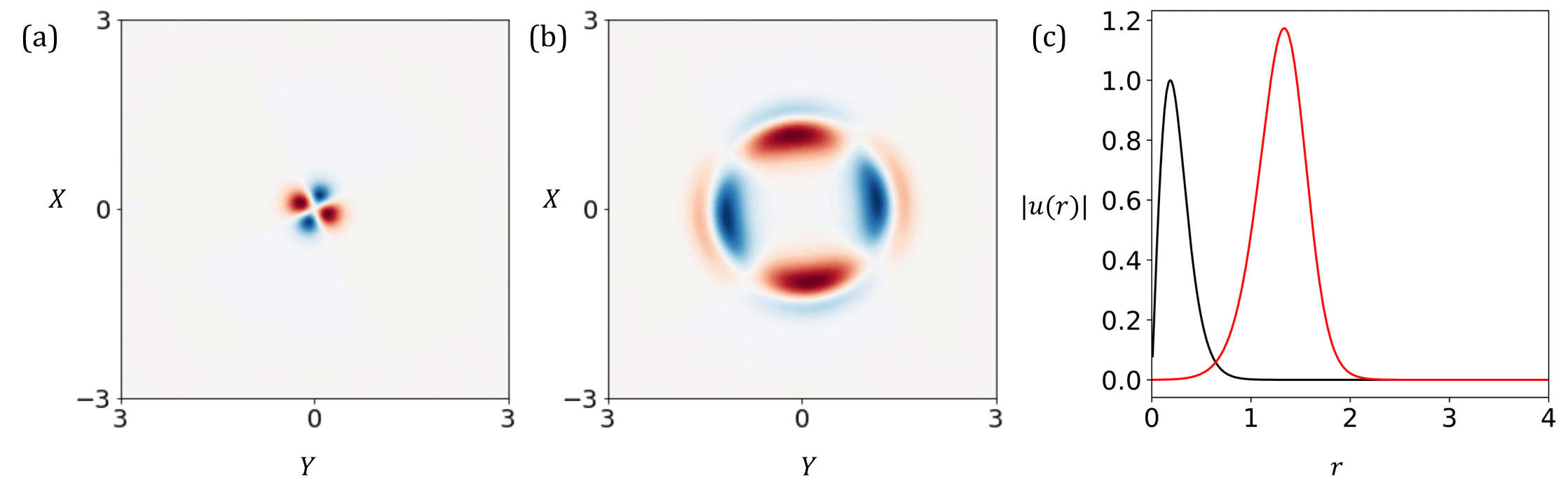}
  \caption{Comparison of the structures of a regular core mode and a regular ring mode,  obtained at $Re = 10^4$, $m = 2$, $kW_0 = -4$. (a) Core mode: 2-D contours of axial disturbance vorticity for the mode with $k = 10.0$, $\Real(\omega) = -3.252$, $\Imag(\omega) = -0.016$; (b) ring mode: $k = 10.0$, $\Real(\omega) = 0.719$, $\Imag(\omega) = -0.011$; (c) radial distribution of the absolute value of the eigenfunction $|u(r)|$, with the black curve corresponding to the mode in (a) and the red curve to the mode in (b).}
\label{fig:batchelormodes_corevsring}
\end{figure}

Based on the radial location where the modes are localized, they can be classified as either core modes or ring modes. A mode is termed a core mode if the vortex centre ($r = 0$) lies within the region where the mode is localized—that is, the mode is confined between the centre and a turning point defined by either of the curves $\omega^{\pm}(r)$. All the modes in the blue and red shaded regions of figure~\ref{fig:batchelormodes_asymcurves}(a) and (b) fall into this category. In contrast, a ring mode is localized away from the vortex centre and lies between two turning points given by $\omega^{\pm}(r)$. The yellow shaded region in figure~\ref{fig:batchelormodes_asymcurves}(b) shows such modes. In general, core modes in a Batchelor vortex are expected for frequencies satisfying $-2 < \omega - m - kW_0 < 2$, whereas ring modes appear outside this range.

Figure~\ref{fig:batchelormodes_corevsring} compares the structures of a core mode and a ring mode, selected respectively from the blue and yellow regions of figure~\ref{fig:batchelormodes_asymcurves}(b). For the ring mode, the region where it is localized is shifted to larger values of $r$, in contrast to the core mode, where it is localized close to the centre. The hollow centre and annular structure characteristic of a ring mode are also clearly visible in its 2-D profile.

\subsubsection*{Predicted resonance domain of triangular instability}
\label{subsubsec:predres}

\begin{figure}
  \centering
  \includegraphics[width=0.6\textwidth]{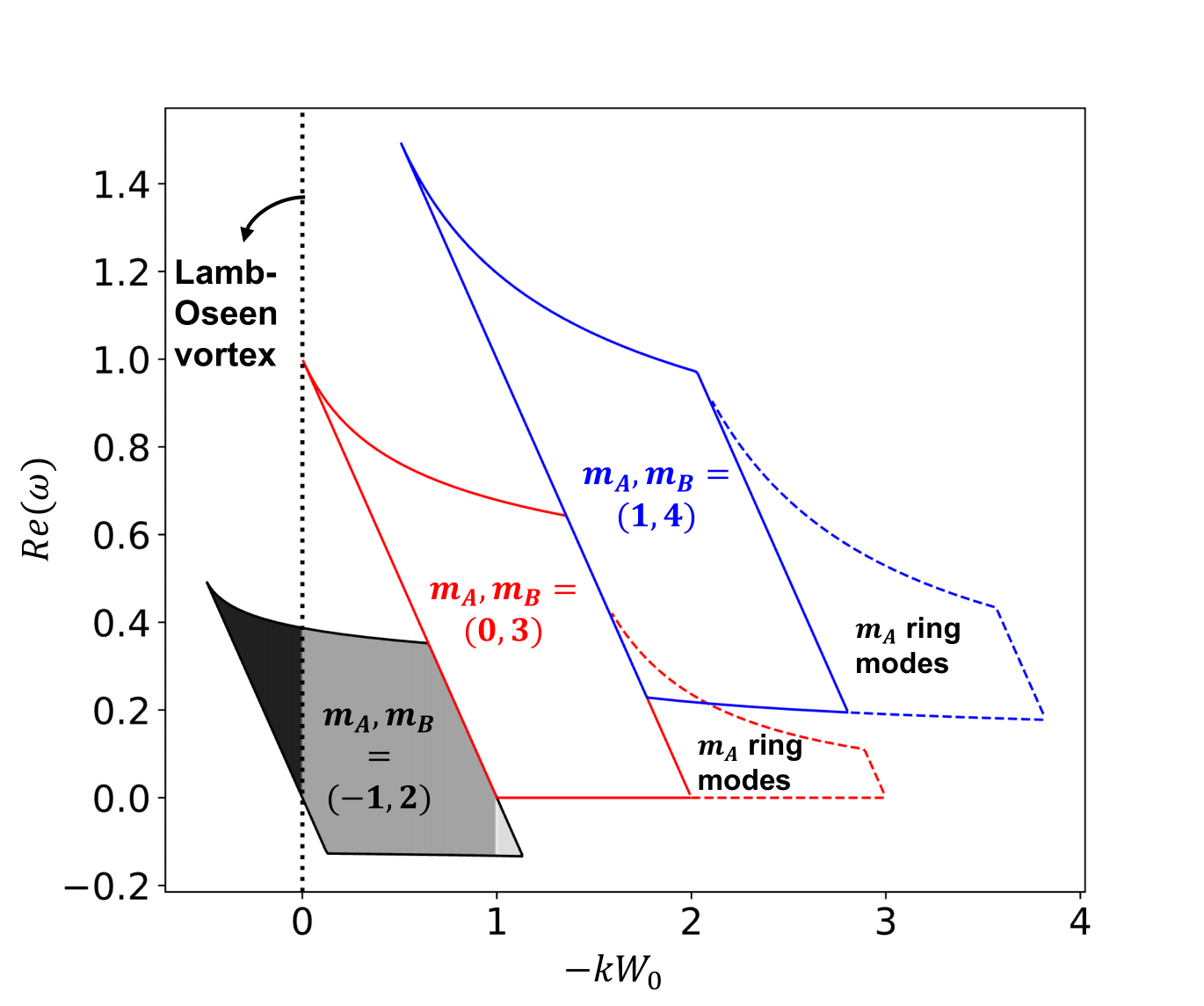}
 \caption{Domain of possible resonance between Kelvin modes of the Batchelor vortex, shown for the azimuthal wavenumber pairs $(m_A, m_B) = (-1, 2)$, $(0, 3)$, and $(1, 4)$, indicated in black, red, and blue, respectively. Solid lines represent resonance domains involving core modes, while dashed lines correspond to ring modes of $m_A$. In the $(-1,2)$ domain, the black shaded region indicates resonances involving singular modes of $m = 2$ only; the mid-grey region corresponds to resonances involving singular modes of both $m = -1$ and $m = 2$; and the pale grey region corresponds to resonances involving singular modes of $m = -1$ only. The black dotted line at $kW_0 = 0$ represents the Lamb--Oseen vortex case.}

\label{fig:resdom}
\end{figure}

In AHL25, we demonstrated that for the Lamb–Oseen vortex, the $m = 2$ (or $-2$) modes in the resonant pairs correspond to singular neutral core modes, while the $m = 1$ (or $-1$) modes are regular neutral core modes. In contrast, for the Batchelor vortex, as mentioned before, each value of $k$ yields distinct epicyclic and critical frequency profiles, allowing for resonant combinations involving a wider range of mode types, including ring modes. By using the expressions for $\omega^{\pm}(r)$ and $\omega_c(r)$ in equations~\eqref{eqc1} and~\eqref{eqc2}, along with the triangular instability resonance condition given in equation~\eqref{eq23}, we analyzed all possible resonant azimuthal wavenumber pairs within the $(\Real(\omega), kW_0)$ parameter space. The resulting resonance domains are shown in figure~\ref{fig:resdom} for the pairs $(m_A,m_B) = (-1,2)$, $(0,3)$, and $(1,4)$. The results can be summarized as follows:

(a) For the pair $(-1,2)$, the nature of the resonant modes depends on the value of $kW_0$ and can be categorized into the following three regions:
   \begin{enumerate}
     \item For $kW_0 \geq 0$: the $m_A$ modes are regular neutral core modes, while the $m_B$ modes are singular neutral core modes.
     \item For $-1 < kW_0 < 0$: both $m_A$ and $m_B$ modes are singular neutral core modes.
     \item For $kW_0 \leq -1$: the $m_A$ modes are singular neutral core modes, while the $m_B$ modes are regular neutral core modes.
   \end{enumerate}

(b) For all higher azimuthal wavenumber pairs such as $(0,3)$, $(1,4)$, $(2,5)$, etc., the nature of the modes are as follows.
   \begin{enumerate}
     \item The modes corresponding to $m_A$ are always regular neutral core modes or ring modes.
     \item The modes corresponding to $m_B$ are always singular neutral core modes.
   \end{enumerate}

\section{Triangular instability characteristics}
\label{sec:tri_instab_char}

In the previous section, we identified the domains in which resonant Kelvin modes may occur and potentially become unstable through triangular instability, and we used asymptotic analysis to characterize the nature of the modes involved in each resonance. Building on that foundation, this section presents the actual evidence of unstable modes and describes their properties in detail. In section~\ref{subsec:theorymath}, we derive the theoretical expression used to compute the growth rate of a resonant pair of Kelvin modes. We then compare, in section~\ref{subsec:numresults}, the growth rates and structural characteristics of the unstable modes obtained from DNS with those predicted by theory using this expression. Once the agreement between DNS and theory is established, we use the predictive power of the theoretical framework to examine in section~\ref{subsec:theoryresults} how the characteristics of the unstable modes evolve continuously as the axial flow is varied. Such a parametric study would be prohibitively expensive using DNS alone, making the theoretical approach particularly useful.

\subsection{Theoretical expression for the triangular instability growth rate}
\label{subsec:theorymath}

The method for calculating the growth rate linked to the resonant coupling between two Kelvin modes and the triangular strain field follows an asymptotic multiscale analysis in the limit of small $\epsilon$, similar to the approach of \citet{moore1975instability}. The perturbation is modeled as a combination of two normal modes with azimuthal wavenumbers $m_A$ and $m_B$ (related through equation~\eqref{eq23}), forming a resonant pair of Kelvin modes at leading order in $\epsilon$. We express the disturbance as
\begin{equation}
 \mathbf{u} \sim \left(A\mathbf{\Tilde{u}}_A(r)e^{\mathrm{i}(m_A\theta)} + B\mathbf{\Tilde{u}}_B(r)e^{\mathrm{i}(m_B\theta)}\right)e^{\mathrm{i}(kz - \omega t)},
\label{eq24}
\end{equation}
where the (real) axial wavenumber $k$ and the (complex) frequency $\omega$ of both modes are assumed to lie near a resonant point defined by $k_c$ and a real frequency $\omega_c$.

Substituting equation~\eqref{eq24} into equation~\eqref{eq20} yields two equations corresponding to the components proportional to $e^{\mathrm{i}m_A\theta}$ and $e^{\mathrm{i}m_B\theta}$, respectively, i.e.
\begin{align}
   A\left( \omega\mathcal{L} - k\mathcal{P} + \mathrm{i}\mathcal{M}(m_A)-\frac{\mathrm{i}}{\Rey}\mathcal{V}(m_A,k)\right)\mathbf{\Tilde{u}}_A &= \mathrm{i}B\epsilon\overline{\mathcal{N}}(m_B)\mathbf{\Tilde{u}}_B, \label{eq25} \\
   B\left( \omega\mathcal{L} - k\mathcal{P} + \mathrm{i}\mathcal{M}(m_B)-\frac{\mathrm{i}}{\Rey}\mathcal{V}(m_B,k)\right)\mathbf{\Tilde{u}}_B &= \mathrm{i}A\epsilon\mathcal{N}(m_A)\mathbf{\Tilde{u}}_A, \label{eq26}
\end{align}
where $\overline{\mathcal{N}}(m_B)$ and $\mathcal{N}(m_A)$ are obtained by substituting $\partial/\partial\theta$ with $\mathrm{i}m_B$ and $\mathrm{i}m_A$, respectively, in the $\mathcal{N}$ and $\overline{\mathcal{N}}$ operators defined in appendix~\ref{appA}. These coupled equations illustrate how the two modes interact via the straining field, represented by the right-hand side terms.

The complex frequency $\omega$ is determined by applying an orthogonality condition using the adjoint resonant Kelvin modes. The adjoint eigenfunctions $\mathbf{\Tilde{u}}_A^{\dagger}$ and $\mathbf{\Tilde{u}}_B^{\dagger}$ are solutions of the adjoint form of equation~\eqref{eq22} for $(m,k) = (m_A,k_c)$ and $(m_B,k_c)$, respectively, defined with respect to the scalar product
\begin{equation}
\left\langle \mathbf{u_1},\mathbf{u_2} \right\rangle = \int_{0}^{\infty}\mathbf{u^*_1}(r)\mathbf{u_2}(r)\, r \, dr,
\label{eq27}
\end{equation}
where `$*$' denotes complex conjugation. We denote the inviscid forms of the Kelvin modes as $(\mathbf{\Tilde{u}}_A^{(\infty)}, \mathbf{\Tilde{u}}_B^{(\infty)})$ and their adjoints as $(\mathbf{\Tilde{u}}_A^{\dagger(\infty)}, \mathbf{\Tilde{u}}_B^{\dagger(\infty)})$. By taking the scalar product of equation \eqref{eq25} with $\mathbf{\Tilde{u}}_A^{\dagger(\infty)}$ and of equation \eqref{eq26} with $\mathbf{\Tilde{u}}_B^{\dagger(\infty)}$, we obtain:
{\begin{align}
   \left( \omega - \omega_A^{(\infty)} - Q_A^{(\infty)}(k - k_c^{(\infty)}) - \mathrm{i}\frac{V_A^{(\infty)}}{\Rey} \right) A = \mathrm{i}\epsilon C_{AB}^{(\infty)} B, \label{eq28} \\
   \left( \omega - \omega_B^{(\infty)} - Q_B^{(\infty)}(k - k_c^{(\infty)}) - \mathrm{i}\frac{V_B^{(\infty)}}{\Rey} \right) B = \mathrm{i}\epsilon C_{BA}^{(\infty)} A, \label{eq29}
\end{align}}
where the coefficients $Q_A^{(\infty)}$, $Q_B^{(\infty)}$, $V_A^{(\infty)}$, $V_B^{(\infty)}$, $C_{AB}^{(\infty)}$, $C_{BA}^{(\infty)}$ are given by
\begin{equation}
Q_A^{(\infty)} = \frac{\left\langle \mathbf{\Tilde{u}}_A^{\dagger(\infty)},\mathcal{P}\mathbf{\Tilde{u}}_A^{(\infty)} \right\rangle}{\left\langle \mathbf{\Tilde{u}}_A^{\dagger(\infty)}\mathcal{L}\mathbf{\Tilde{u}}_A^{(\infty)} \right\rangle}, \;\;\;Q_B^{(\infty)} = \frac{\left\langle \mathbf{\Tilde{u}}_B^{\dagger(\infty)},\mathcal{P}\mathbf{\Tilde{u}}_B^{(\infty)} \right\rangle}{\left\langle \mathbf{\Tilde{u}}_B^{\dagger(\infty)},\mathcal{L}\mathbf{\Tilde{u}}_B^{(\infty)} \right\rangle},
\label{eq30}
\end{equation}

\begin{equation}
V_A^{(\infty)}= \frac{\left\langle \mathbf{\Tilde{u}}_A^{\dagger(\infty)},\mathcal{V}\mathbf{\Tilde{u}}_A^{(\infty)} \right\rangle}{\left\langle \mathbf{\Tilde{u}}_A^{\dagger(\infty)},\mathcal{L}\mathbf{\Tilde{u}}_A^{(\infty)}\right\rangle}, \;\;\;V_B^{(\infty)} = \frac{\left\langle \mathbf{\Tilde{u}}_B^{\dagger(\infty)},\mathcal{V}\mathbf{\Tilde{u}}_B^{(\infty)} \right\rangle}{\left\langle \mathbf{\Tilde{u}}_B^{\dagger(\infty)},\mathcal{L}\mathbf{\Tilde{u}}_B^{(\infty)} \right\rangle},
\label{eq31}
\end{equation}

\begin{equation}
C_{AB}^{(\infty)} = \frac{\left\langle \mathbf{\Tilde{u}}_A^{\dagger},\overline{\mathcal{N}}(m_B)\mathbf{\Tilde{u}}_B^{(\infty)} \right\rangle}{\left\langle \mathbf{\Tilde{u}}_A^{\dagger(\infty)},\mathcal{L}\mathbf{\Tilde{u}}_A^{(\infty)} \right\rangle}, \;\;\;C_{BA}^{(\infty)} = \frac{\left\langle \mathbf{\Tilde{u}}_B^{\dagger(\infty)},\mathcal{N}(m_A)\mathbf{\Tilde{u}}_A^{(\infty)} \right\rangle}{\left\langle \mathbf{\Tilde{u}}_B^{\dagger(\infty)},\mathcal{L}\mathbf{\Tilde{u}}_B^{(\infty)}\right\rangle}
\label{eq32}.
\end{equation}

The frequencies $\omega_A^{(\infty)}$ and $\omega_B^{(\infty)}$ represent the inviscid estimates of the resonant Kelvin mode frequencies at $k = k_c^{(\infty)}$. They can be written as
\begin{equation} 
\omega_A^{(\infty)} = \omega_c^{(\infty)} + \mathrm{i} \Imag\left(\omega_A^{(\infty)}\right), \quad
\omega_B^{(\infty)} = \omega_c^{(\infty)} + \mathrm{i} \Imag\left(\omega_B^{(\infty)}\right),
\label{eq33}
\end{equation}
where $\Imag\left(\omega_A^{(\infty)}\right)$ and $\Imag\left(\omega_B^{(\infty)}\right)$ are the critical-layer damping rates of the respective modes.

Equations \eqref{eq28}-\eqref{eq29} with \eqref{eq33} give a quadratic equation for the complex frequency $\omega$ as a function of $k$, i.e.
\begin{align}
\left(  \omega - \omega_A^{(\infty)} - Q_A^{(\infty)}(k - k_c^{(\infty)}) - \mathrm{i}\frac{V_A^{(\infty)}}{\Rey} \right) &\left( \omega - \omega_B^{(\infty)} - Q_B^{(\infty)}(k - k_c^{(\infty)}) - \mathrm{i}\frac{V_B^{(\infty)}}{\Rey} \right) \nonumber \\
&= -\epsilon^2 (N^{(\infty)})^2,
\label{eq34Gen}
\end{align}
where
\begin{equation}
N^{(\infty)} = \sqrt{C_{AB}^{(\infty)}C_{BA}^{(\infty)}}.
\label{eq34sub}
\end{equation}

The growth rate $\sigma$ is defined as $\Imag(\omega)$, and a growth rate band around the resonant axial wavenumber $k_c^{(\infty)}$ can be computed using the equations above. Both modes oscillate with a common resonant frequency given by $\Real(\omega)$. The right-hand side represents the coupling term driving the instability. The terms $\mathrm{i}V_A^{(\infty)}/\Rey$ and $\mathrm{i}V_B^{(\infty)}/\Rey$ are volumic viscous damping terms and vanish when computing the inviscid growth rate. The following equations:
\begin{align}
\omega - \omega_A^{(\infty)} - Q_A^{(\infty)}(k - k_c^{(\infty)}) = 0, \\ 
\omega - \omega_B^{(\infty)} - Q_B^{(\infty)}(k - k_c^{(\infty)}) = 0,
\label{eq35}
\end{align}
are the linear approximates of the inviscid dispersion relations for the respective azimuthal wavenumbers of the unstrained vortex. These provide the value of $\omega$ for modes with $k = k_c^{(\infty)} + \tau$ near the resonant point and are expected to closely approximate the original dispersion relations for small $\tau$ up to $O(10^{-1})$. In particular, the linear approximation of the variation of critical-layer damping values with $k$ can therefore be expressed as
\begin{align}
\omega_{\text{CL},A}^{(\infty)} = \Imag(\omega_A^{(\infty)}) + \Imag(Q_A^{(\infty)})\tau, \label{eq36a} \\ 
\omega_{\text{CL},B}^{(\infty)} = \Imag(\omega_B^{(\infty)}) + \Imag(Q_B^{(\infty)})\tau,
\label{eq36b}
\end{align}
respectively, for each mode. If $\tau$ becomes too large, the linear approximation breaks down and the predicted critical-layer damping may incorrectly become positive, which is non-physical. While this could, in principle, be corrected by retaining higher-order terms in $\tau$, such refinements lie beyond the scope of the present analysis. Instead, for all $k = k_c^{(\infty)} + \tau$, any positive values of critical-layer damping obtained from equations~\eqref{eq36a} and \eqref{eq36b} are set to zero when computing the growth rate. This adjustment does not affect the behaviour at the resonant point $k_c$, and is required only for values of $k$ sufficiently far from $k_c$ where the linear approximation ceases to be valid. This procedure is applied consistently throughout the following two sections.

\subsection{Unstable Modes: Numerical Results and Theoretical Predictions}
\label{subsec:numresults}

\begin{figure}
  \centering
  \includegraphics[width=0.8\textwidth]{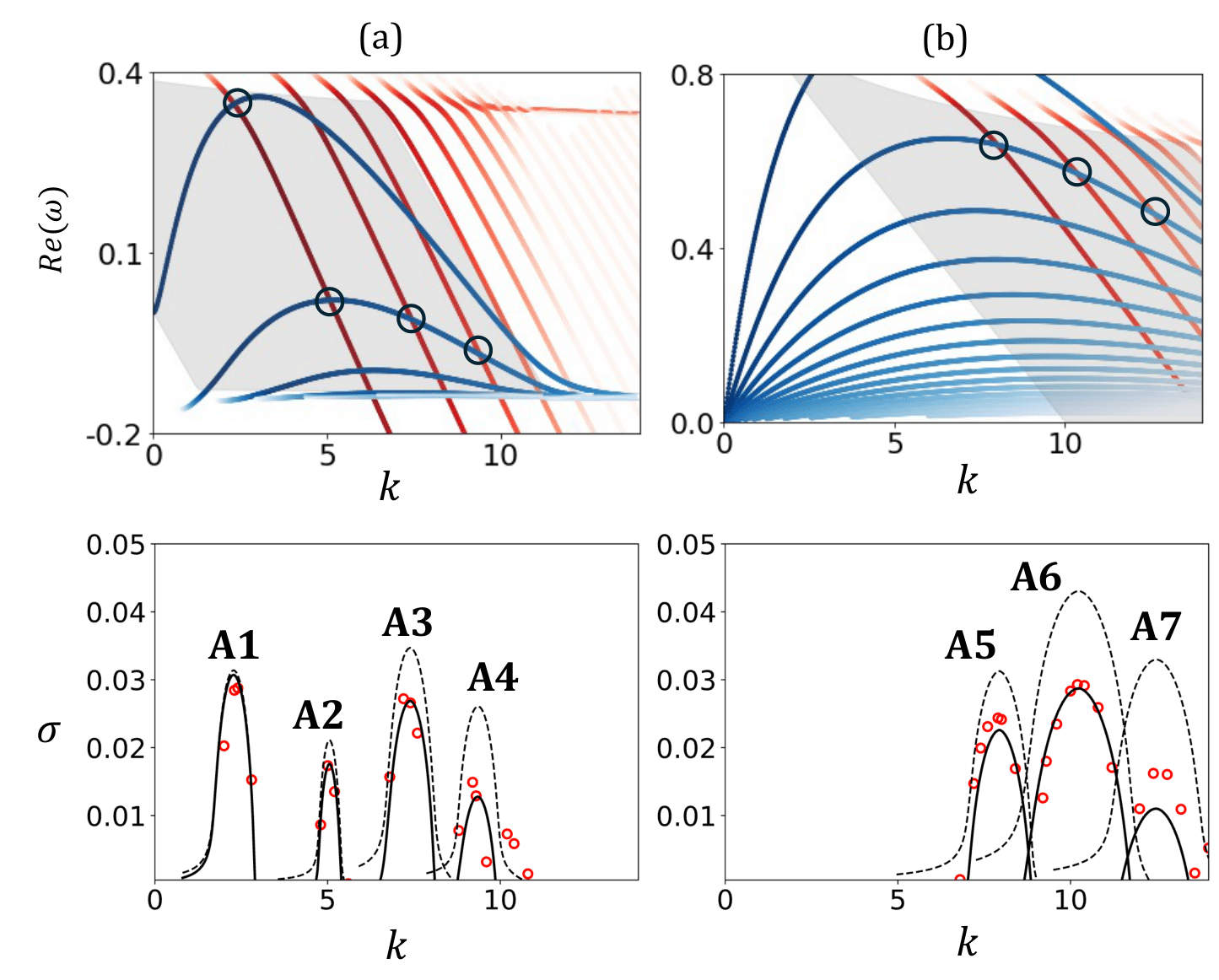}
  \caption{Resonant modes A1, A2, etc. and their growth rates at $W_0 = -0.1$;  Top panels: dispersion curves showing $\Real(\omega)$ versus $k$ for (a) $(m_A, m_B) = (-1, 2)$ and (b) $(0, 3)$. Red curves represent $m_B$ modes, blue curves represent $m_A$ modes, with color transitions indicating increasing damping (from 0 to $-0.05$). Resonant modes  are marked with black circles at the crossing points. Grey background regions in the top panels mark the core mode resonance domain predicted by the asymptotic theory. Bottom panels: growth rate curves $\sigma$ versus $k$; solid black lines show the viscous case ($\Rey = 10^4$), dashed lines show the inviscid case, and red circles indicate DNS results.}

\label{fig:res_gw_curves_w1}
\end{figure}

\begin{figure}
  \centering
  \includegraphics[width=1.0\textwidth]{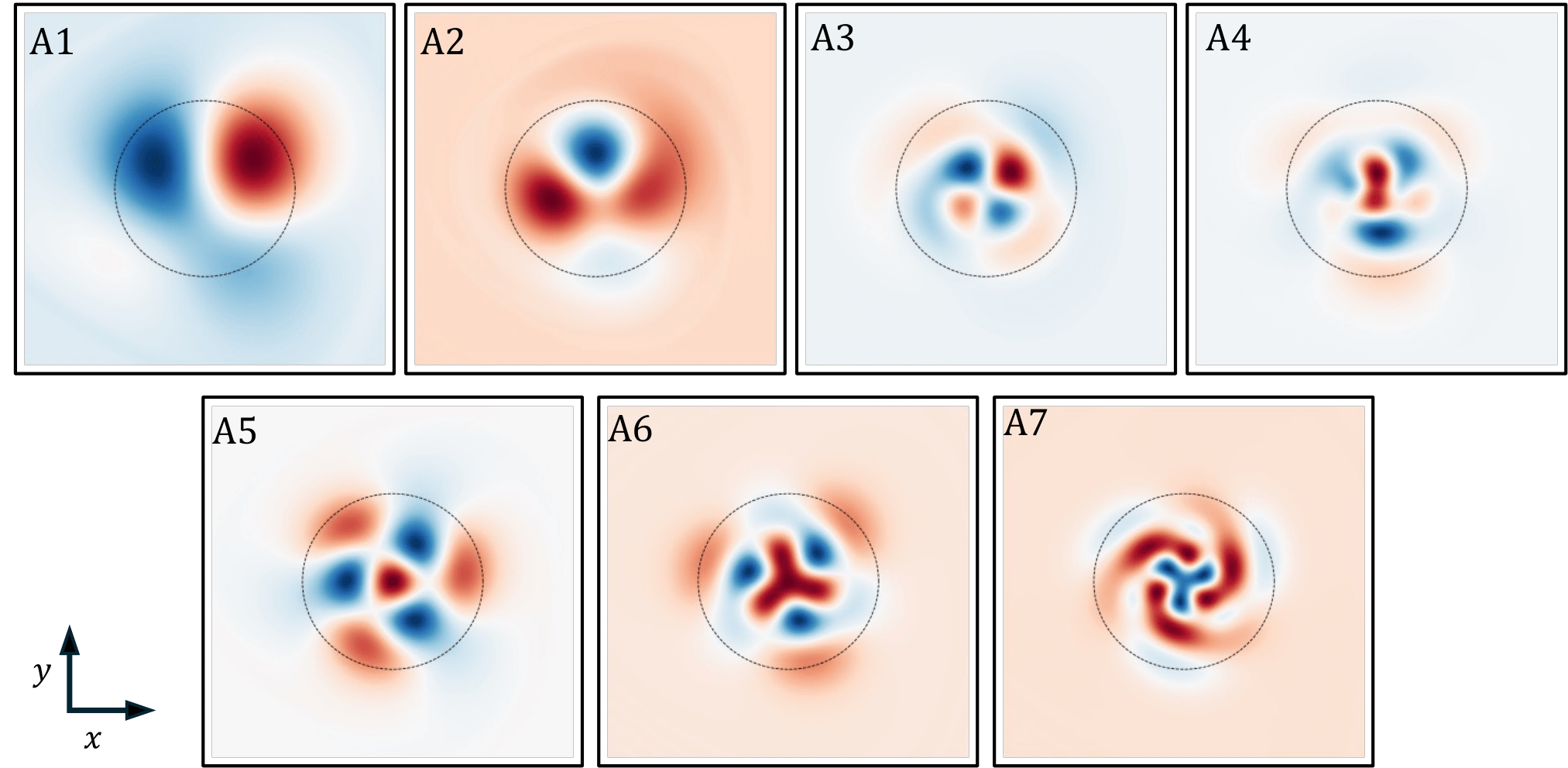}
  \caption{Axial flow strength $W_0 = -0.1$; 2-D structures of the modes labeled A1, A2, etc., obtained from DNS, showing contours of axial vorticity in the $x$–$y$ plane. Red regions indicate positive vorticity, while blue regions indicate negative vorticity. Domain: $-2 < X,Y < 2$. The hub vortex boundary is indicated with a dotted circle for reference.}
\label{fig:2dstructures1}
\end{figure}

\begin{figure}
  \centering
  \includegraphics[width=1.0\textwidth]{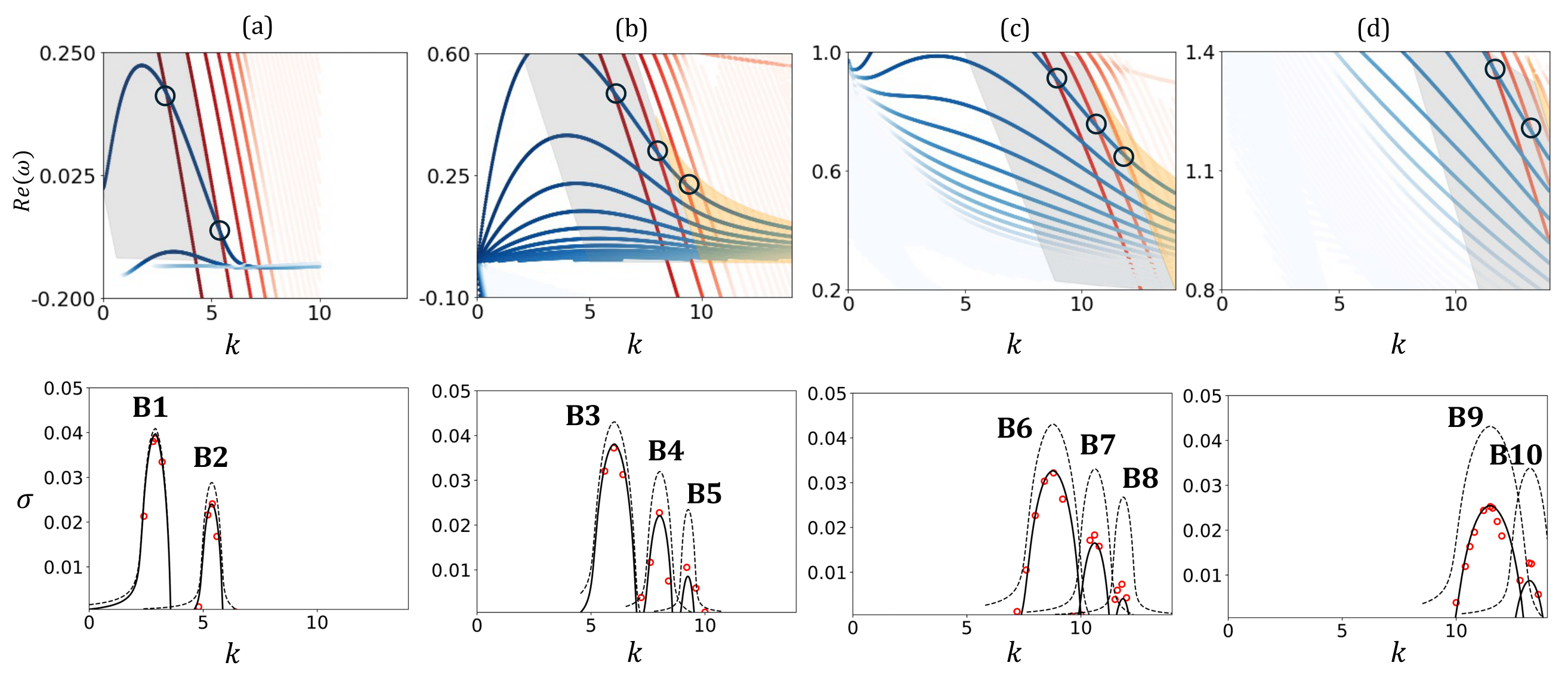}
  \caption{Resonant modes B1, B2, etc. and their growth rates at $W_0 = -0.2$; Top panels: dispersion curves showing $\Real(\omega)$ versus $k$ for (a) $(m_A, m_B) = (-1, 2)$, (b) $(0, 3)$, (c) $(1, 4)$, and (d) $(2, 5)$. Red curves represent $m_B$ modes, blue curves represent $m_A$ modes, with color transitions indicating increasing damping (from 0 to $-0.05$). Resonant modes are marked with black circles at the crossing points. Grey background regions in the top panels mark the core mode resonance domain predicted by the asymptotic theory, while orange regions represent the resonance domain for ring modes of $m_A$. Bottom panels: growth rate curves $\sigma$ versus $k$; solid black lines show the viscous case ($\Rey = 10^4$), dashed lines show the inviscid case, and red circles indicate DNS results. }
\label{fig:res_gw_curves_w2}
\end{figure}

\begin{figure}
  \centering
  \includegraphics[width=1.0\textwidth]{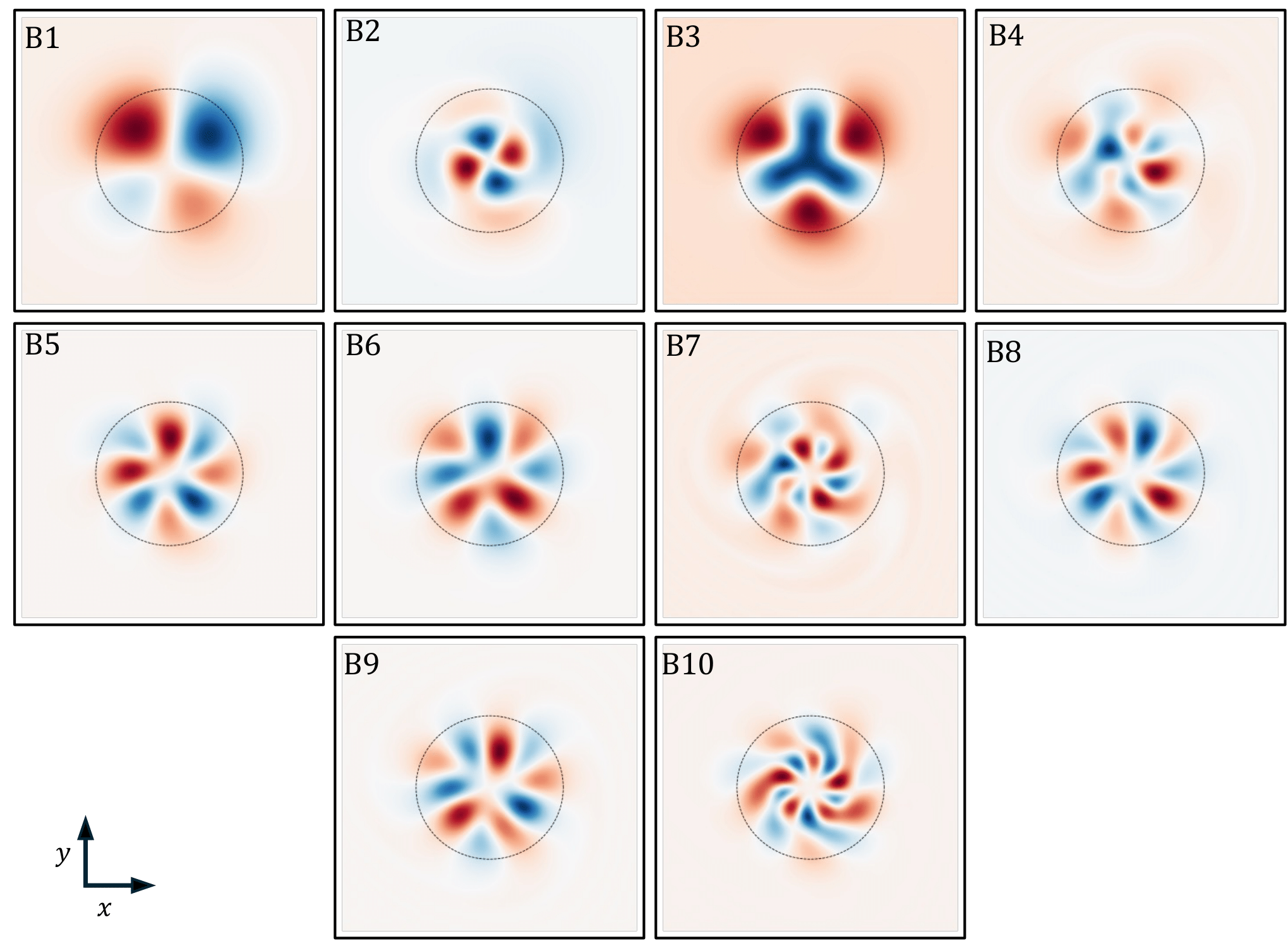}
  \caption{Axial flow strength $W_0 = -0.2$; 2-D structures of the modes labeled B1, B2, etc., obtained from DNS, showing contours of axial vorticity in the $x$–$y$ plane. Red regions indicate positive vorticity, while blue regions indicate negative vorticity. Domain: $-2 < X,Y < 2$. The hub vortex boundary is indicated with a dotted circle for reference.}
\label{fig:2dstructures2}
\end{figure}

\begin{figure}
  \centering
  \includegraphics[width=1.0\textwidth]{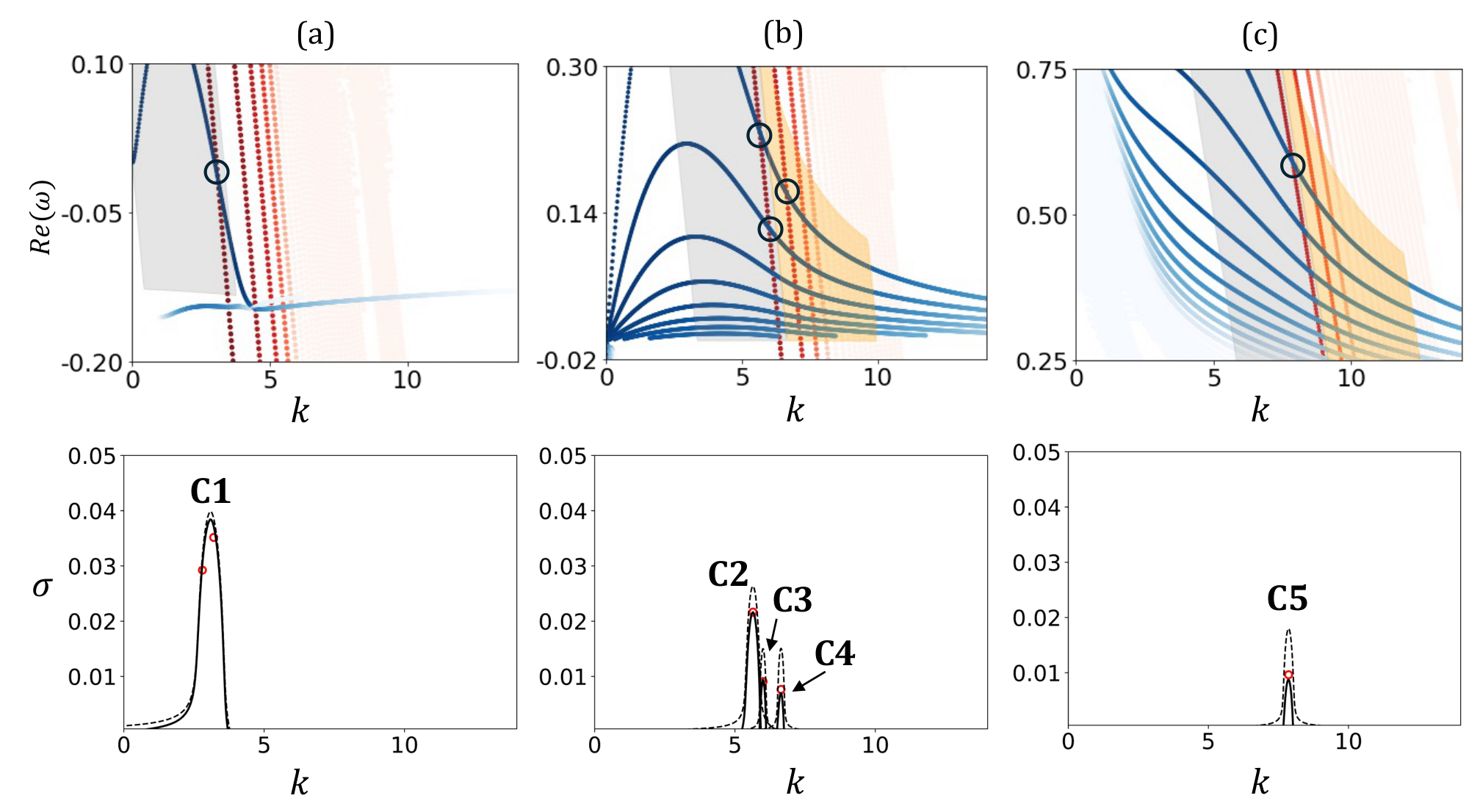}
  \caption{Resonant modes C1, C2, etc. and their growth rates at $W_0 = -0.3$; Top panels: dispersion curves showing $\Real(\omega)$ versus $k$ for (a) $(m_A, m_B) = (-1, 2)$, (b) $(0, 3)$, and (c) $(1, 4)$. Blue–green curves represent $m_B$ modes, black–brown curves represent $m_A$ modes, with color transitions indicating increasing damping (from 0 to $-0.05$). Resonant modes are marked with black circles at the crossing points. Grey background regions in the top panels mark the core mode resonance domain predicted by the asymptotic theory, while orange regions represent the resonance domain for ring modes of $m_A$. Bottom panels: growth rate curves $\sigma$ versus $k$; solid black lines show the viscous case ($\Rey = 10^4$), dashed lines show the inviscid case, and red circles indicate DNS results.}
\label{fig:res_gw_curves_w3}
\end{figure}

\begin{figure}
  \centering
  \includegraphics[width=0.8\textwidth]{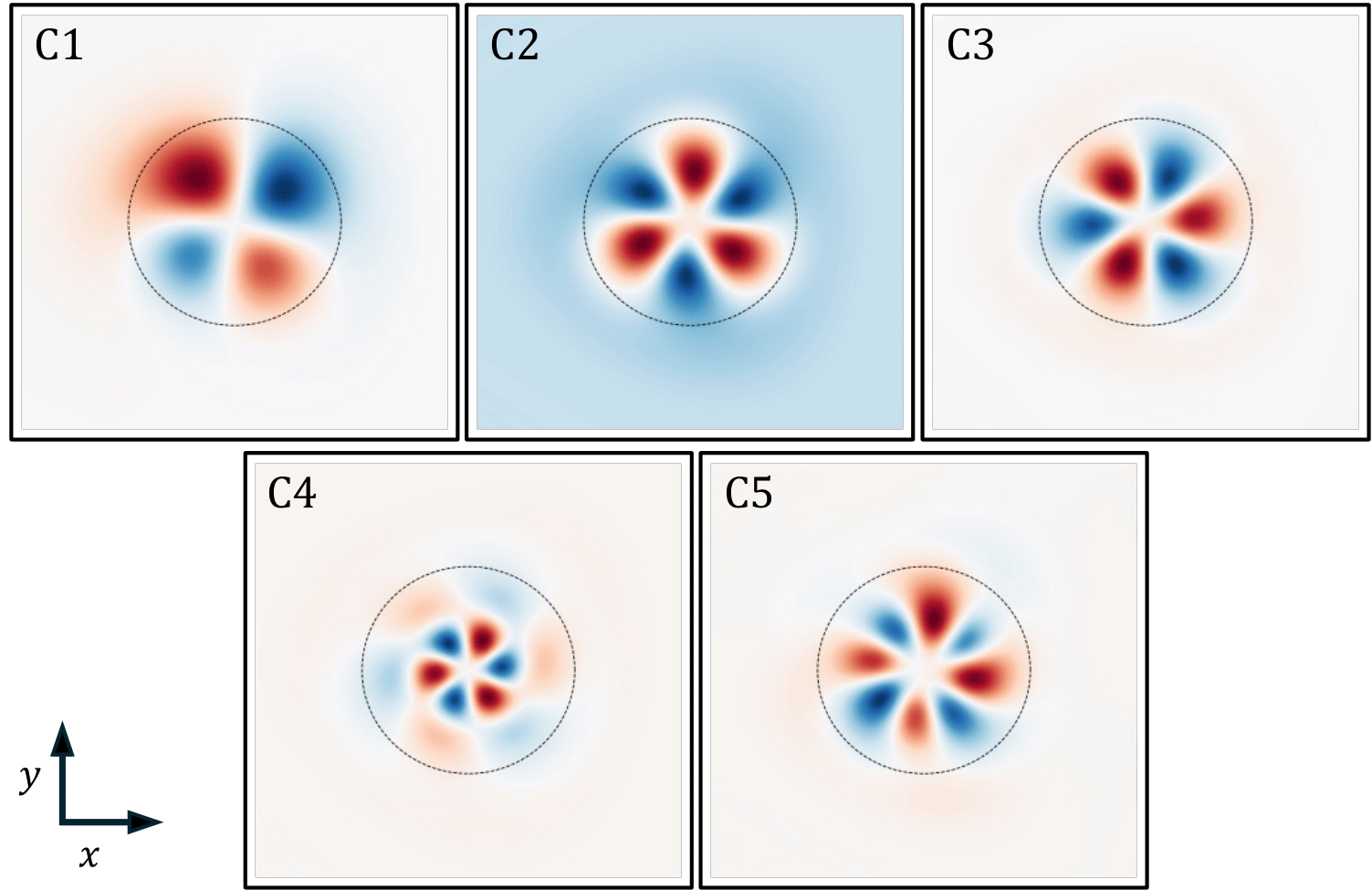}
  \caption{Axial flow strength $W_0 = -0.3$; 2-D structures of the modes labeled C1, C2, etc., obtained from DNS, showing contours of axial vorticity in the $x$–$y$ plane. Red regions indicate positive vorticity, while blue regions indicate negative vorticity. Domain: $-2 < X,Y < 2$. The hub vortex boundary is indicated with a dotted circle for reference.}
\label{fig:2dstructures3}
\end{figure}

\begin{table}
\centering
\begin{tabular}{c c c c c c c c c c c}
   $W_0$ & Mode label & \multicolumn{2}{c}{$(m_A,m_B,[l_A,l_B])$} & $m=-1$ & $0$ & $1$ & $2$ & $3$ & $4$ & $5$ \\
    &  & Primary & Secondary &  &  &  &  &  &  &  \\
\hline
-0.1 & A1 & $(-1,2,[1,1])$ & -- & 0.766 & 0 & 0 & 0.233 & 0 & 0 & 0 \\
     & A2 & $(-1,2,[2,1])$ & -- & 0.403 & 0.010 & 0 & 0.584 & 0 & 0 & 0 \\
     & A3 & $(-1,2,[2,2])$ & -- & 0.470 & 0.006 & 0 & 0.518 & 0.005 & 0 & 0 \\
     & A4 & $(-1,2,[2,3])$ & A6 & 0 & 0.508 & 0 & 0 & 0.491 & 0 & 0 \\
     & A5 & $(0,3,[2,1])$ & A3 & 0.110 & 0.433 & 0 & 0.105 & 0.351 & 0 & 0 \\
     & A6 & $(0,3,[2,2])$ & -- & 0 & 0.538 & 0 & 0 & 0.461 & 0 & 0 \\
     & A7 & $(0,3,[2,3])$ & -- & 0 & 0.551 & 0 & 0 & 0.448 & 0 & 0 \\
\hline
-0.2 & B1  & $(-1,2,[1,1])$ & -- & 0.621 & 0 & 0 & 0.379 & 0 & 0 & 0 \\
     & B2  & $(-1,2,[1,2])$ & -- & 0.560 & 0 & 0 & 0.440 & 0 & 0 & 0 \\
     & B3  & $(0,3,[1,1])$  & -- & 0 & 0.545 & 0 & 0 & 0.455 & 0 & 0 \\
     & B4  & $(0,3,[1,2])$  & B6 & 0 & 0.197 & 0.371 & 0 & 0.175 & 0.256 & 0 \\
     & B5  & $(0,3,[1,3])$  & B6 & 0 & 0 & 0.531 & 0 & 0 & 0.468 & 0 \\
     & B6  & $(1,4,[1,1])$  & -- & 0 & 0 & 0.513 & 0 & 0 & 0.486 & 0 \\
     & B7  & $(1,4,[1,2])$  & B9 & 0 & 0 & 0.380 & 0.174 & 0 & 0.330 & 0.114 \\
     & B8  & $(1,4,[1,3])$  & B9 & 0 & 0 & 0 & 0.535 & 0 & 0 & 0.464 \\
     & B9  & $(2,5,[1,1])$  & -- & 0 & 0 & 0 & 0.521 & 0 & 0 & 0.478 \\
     & B10 & $(2,5,[1,2])$  & -- & 0 & 0 & 0 & 0.541 & 0 & 0 & 0.457 \\
\hline
-0.3 & C1 & $(-1,2,[1,1])$ & -- & 0.554 & 0 & 0 & 0.446 & 0 & 0 & 0 \\
     & C2 & $(0,3,[1,1])$  & -- & 0 & 0.429 & 0 & 0 & 0.570 & 0 & 0 \\
     & C3 & $(0,3,[2,1])$  & -- & 0.001 & 0.285 & 0.017 & 0.004 & 0.692 & 0 & 0 \\
     & C4 & $(0,3,[1,2])$  & -- & 0 & 0.427 & 0.009 & 0.002 & 0.561 & 0 & 0 \\
     & C5 & $(1,4,[1,1])$  & -- & 0 & 0 & 0.445 & 0 & 0 & 0.554 & 0 \\
\end{tabular}

\caption{Composition of the numerically obtained modes for $W_0 = -0.1$, $-0.2$, and $-0.3$. 
The second column lists the mode labels (A, B, C) corresponding to the respective $W_0$ values. 
The third and fourth columns indicate the primary azimuthal wavenumber pairs $(m_A, m_B)$ and their corresponding branches $(l_A, l_B)$, and any secondary resonant pairs, if present. 
Columns 5–11 show the energy distribution among different azimuthal wavenumbers $m$, representing the relative contribution of each $m$-component to the total mode energy.}
\label{tab:modecomp_all}
\end{table}

\begin{table}
\centering
\begin{tabular}{c c c c c c c c}
\hline
$W_0$ & Mode label & $k$ & $\omega_{\text{dns}}$ & $\omega_{\text{th}}$ & $\sigma_{\text{dns}}$ & $\sigma_{\text{th}}$ & $\sigma_{\text{th}}^{(\infty)}$ \\
\hline
-0.1 & A1 & 2.30 & 0.349 & 0.347 & 0.0288 & 0.0307 & 0.0314 \\
     & A2 & 5.06 & 0.035 & 0.021 & 0.0173 & 0.0176 & 0.0211 \\
     & A3 & 7.40 & 0.002 & -0.010 & 0.0266 & 0.0260 & 0.0347 \\
     & A4 & 9.36 & -0.054 & -0.068 & 0.0129 & 0.0127 & 0.0261 \\
     & A5 & 7.96 & 0.654 & 0.639 & 0.0242 & 0.0226 & 0.0313 \\
     & A6 & 10.26 & 0.586 & 0.575 & 0.0292 & 0.0287 & 0.0431 \\
     & A7 & 12.50 & 0.492 & 0.484 & 0.0160 & 0.0110 & 0.0330 \\
\hline
-0.2 & B1 & 2.92 & 0.181 & 0.177 & 0.0387 & 0.0397 & 0.0409 \\
     & B2 & 5.38 & -0.067 & -0.076 & 0.0241 & 0.0240 & 0.0288 \\
     & B3 & 6.04 & 0.512 & 0.494 & 0.0372 & 0.0381 & 0.0431 \\
     & B4 & 8.02 & 0.334 & 0.316 & 0.0227 & 0.0221 & 0.0319 \\
     & B5 & 9.26 & 0.251 & 0.226 & 0.0105 & 0.0090 & 0.0235 \\
     & B6 & 8.80 & 0.942 & 0.921 & 0.0321 & 0.0326 & 0.0431 \\
     & B7 & 10.62 & 0.774 & 0.751 & 0.0183 & 0.0165 & 0.0331 \\
     & B8 & 11.86 & 0.677 & 0.650 & 0.0073 & 0.0041 & 0.0267 \\
     & B9 & 11.52 & 1.395 & 1.367 & 0.0252 & 0.0254 & 0.0431 \\
     & B10 & 13.24 & 1.227 & 1.198 & 0.0127 & 0.0087 & 0.0338 \\
\hline
-0.3 & C1 & 3.10 & -0.029 & -0.014 & 0.0376 & 0.0384 & 0.0398 \\
     & C2 & 5.64 & 0.246 & 0.230 & 0.0216 & 0.0216 & 0.0263 \\
     & C3 & 6.00 & 0.141 & 0.123 & 0.0091 & 0.0094 & 0.0150 \\
     & C4 & 6.64 & 0.178 & 0.161 & 0.0076 & 0.0070 & 0.0151 \\
     & C5 & 7.86 & 0.612 & 0.592 & 0.0096 & 0.0087 & 0.0179 \\
\hline
\end{tabular}

\caption{Summary of the primary resonant modes obtained for $W_0 = -0.1$, $-0.2$, and $-0.3$. 
For each mode, the table lists the axial wavenumber $k$, frequency $\omega_{\text{dns}}$ obtained from DNS, theoretical frequency $\omega_{\text{th}}$, growth rate $\sigma_{\text{dns}}$ from DNS at $\Rey = 10^4$, viscous growth rate $\sigma_{\text{th}}$ from theory at $\Rey = 10^4$, and inviscid growth rate $\sigma_{\text{th}}^{(\infty)}$ from theory.}
\label{tab:sigmafreq_all}
\end{table}

\begin{figure}
  \centering
  \includegraphics[width=1.0\textwidth]{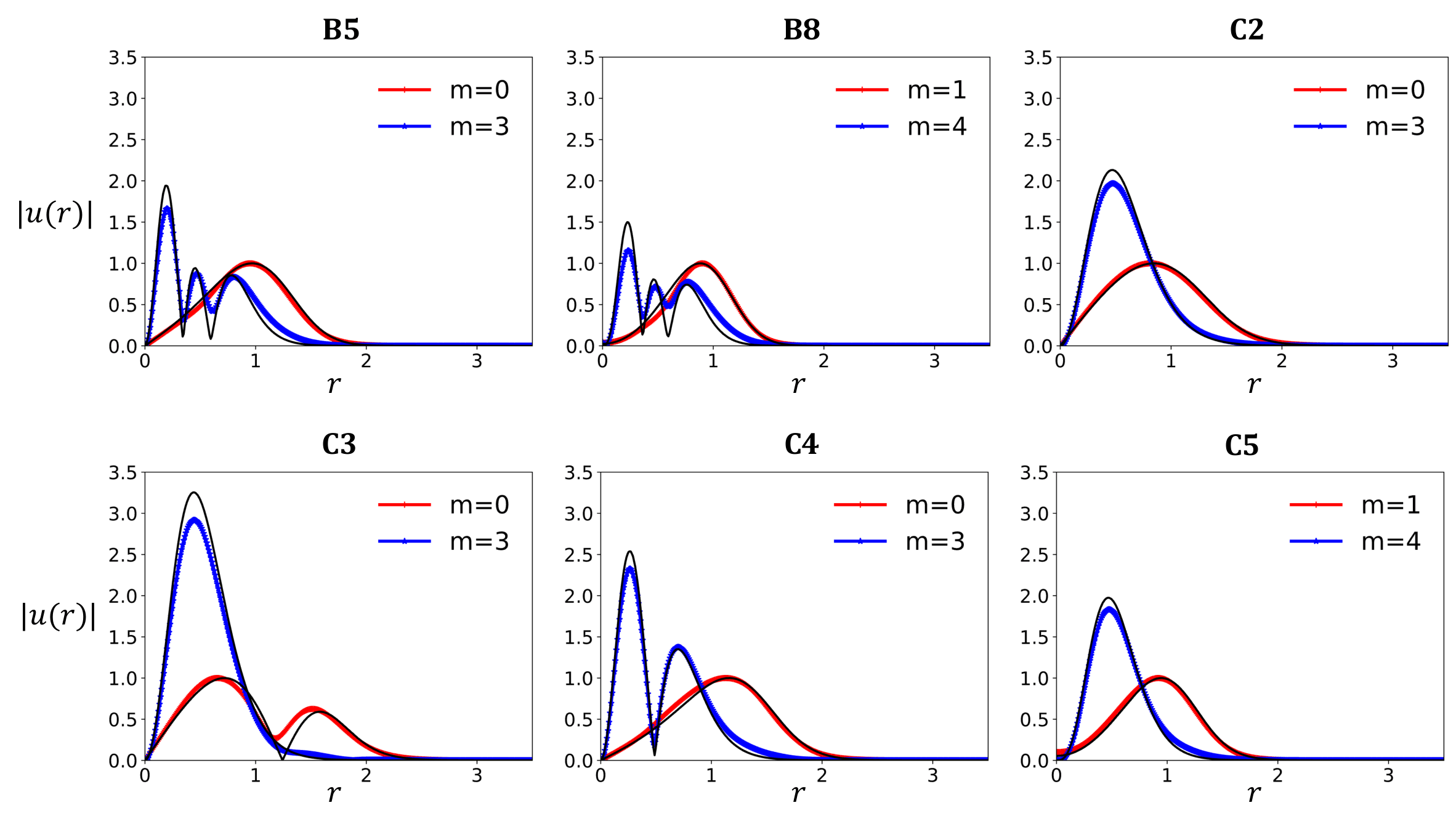}
  \caption{Distributions of the eigenfunctions $|u(r)|$ versus $r$ for modes B5, B8, C2, C3, C4, and C5, where the modes corresponding to $m_A$ are ring modes. Colored lines represent DNS results, while black lines correspond to eigenfunctions obtained by solving equation~\eqref{eq22}.}
\label{fig:ringmode_structures}
\end{figure}

For any resonant pair of Kelvin modes with azimuthal wavenumbers $m_A$ and $m_B$, the theoretical growth rate can be evaluated using equation~\eqref{eq34Gen}. We can thus compare and validate the unstable modes identified numerically through DNS against the predictions of the theoretical framework. To this end, three sets of linearized Navier-Stokes simulations were performed using base flows corresponding to $W_0 = -0.1$, $-0.2$, and $-0.3$ at $\Rey = 10^4$ and remind the reader that $\epsilon = 0.008$ for these base flows. These particular values of $W_0$ were chosen simply because they are small enough to ensure that no inviscid or viscous centre modes of the unstrained Batchelor vortex become unstable, as discussed in the introduction. They also provide a sufficiently broad range of unstable modes to allow meaningful comparison and interpretation. The choice $\Rey = 10^4$ is large enough for non-viscous approximations to be meaningful and sufficiently large for the unstable modes to not be completely suppressed by viscous damping, while also still keeping the DNS computationally tractable. From a practical point of view, although typical wind-turbine Reynolds numbers are much higher, we argue that our study is particularly relevant for laboratory-scale experiments where $\Rey$ is in the range $10^3$–$10^4$.

Starting from a random initial energy distribution, unstable modes with various azimuthal wavenumbers grow exponentially in time. Their growth rate is obtained as half the slope of the logarithmic time evolution of the mode's energy, while the mode frequency is determined by tracking the phase of the disturbance and measuring its angular rotation rate about the $z$-axis. 

For ease of reference, the resonant modes obtained at $W_0 = -0.1$ are labeled A1, A2, etc.; those at $W_0 = -0.2$ as B1, B2, etc.; and those at $W_0 = -0.3$ as C1, C2, etc. For each of these three values of $W_0$, we present both the dispersion and growth-rate curves of the resonant modes, together with their two-dimensional structures obtained from DNS.
For $W_0 = -0.1$, figure~\ref{fig:res_gw_curves_w1} shows, in the top panels, the dispersion curves of the Kelvin modes obtained by numerically solving the set of equations \eqref{eq22} as discussed in section \ref{subsec:kelvinmodes_num}, and the corresponding crossing points for each azimuthal wavenumber pair $(m_A, m_B)$. The resonant modes that yield positive growth rates at $\Rey = 10^4$ are highlighted with circles. In the bottom panels of the same figure, we plot the corresponding theoretical and DNS growth rate curves $\sigma$ as functions of $k$ for the circled modes. These growth rate curves are also compared with the corresponding inviscid growth rate curves predicted by theory. The purpose of comparing with inviscid growth rates is to highlight the maximum growth rate attainable by each mode, which occurs in the inviscid limit. The associated $x$–$y$ plane structures of these unstable modes, shown as contours of the axial disturbance vorticity field, are presented in figure~\ref{fig:2dstructures1}. The periphery of the base hub vortex is indicated in the figure using a dotted circle with radius equal to one for reference. The same organization is used for $W_0 = -0.2$ in figures~\ref{fig:res_gw_curves_w2} and~\ref{fig:2dstructures2}, and for $W_0 = -0.3$ in figures~\ref{fig:res_gw_curves_w3} and~\ref{fig:2dstructures3}.
In all cases, only the circled modes in the top panels exhibit positive growth rates at $\Rey = 10^4$, as they originate from the first few branches (up to the third, at most) of the dispersion curves. Modes associated with higher branches undergo strong volumic viscous damping and/or weak coupling, leading to negligible growth rates even in the inviscid limit. Consequently, our analysis and comparison focus exclusively on these circled modes, labeled according to the notation described above.

From the bottom panels of the figures~\ref{fig:res_gw_curves_w1}, \ref{fig:res_gw_curves_w2} and \ref{fig:res_gw_curves_w3}, we can see that for $W_0 = -0.1$, the mode pairs $(m_A, m_B) = (-1, 2)$ and $(0, 3)$ become resonant and exhibit positive growth rates. At $W_0 = -0.2$, the pairs $(-1, 2)$, $(0, 3)$, $(1, 4)$, and $(2, 5)$ show modes with positive growth rates. At $W_0 = -0.3$, positive growth rates are again obtained for modes from $(-1, 2)$, $(0, 3)$, and $(1, 4)$. Overall, the numerical results and theoretical predictions of the growth rates show good agreement, with only modes A7, B5, B8, and B10 exhibiting slight under-prediction by the theory. Additionally, in the bottom panel of figure~\ref{fig:res_gw_curves_w1}(a), we observe another mode to the right of A4 that shows a positive growth rate in the numerical results, while the theoretical prediction yields zero growth. Upon closer inspection, it appears that unstable modes associated with higher branches tend to exhibit this under-prediction of growth rates by the theory. This is most likely due to the over-estimation of the volumic viscous damping in the theoretical model—the coefficients $V_A^{(\infty)}$ and $V_B^{(\infty)}$—whose magnitudes are generally larger for modes belonging to higher branches (see Table 1 in AHL25, for example). As a result, the growth rates of such modes are more sensitive to even small over-predictions of these viscous damping terms, leading to discrepancies between theory and DNS.

Table~\ref{tab:modecomp_all} summarizes the composition of all numerically obtained modes for $W_0 = -0.1$, $-0.2$, and $-0.3$. For each mode, the table lists the primary azimuthal wavenumber pair $(m_A, m_B, [l_A, l_B])$, any secondary resonant pairs when present, and the corresponding energy distribution among different azimuthal wavenumbers $m$. Note that some of the A, B, and C modes share the same $(m_A, m_B, [l_A, l_B])$ composition, with only their characteristics and structures changing as the axial flow varies. For example, A1, B1, and C1 all correspond to $(-1,2,[1,1])$; B3 and C2 correspond to $(0,3,[1,1])$; A5 and C3 to $(0,3,[2,1])$; and B6 and C5 to $(1,4,[1,1])$.

It is also worth noting that, upon examining the growth-rate curves (and the dispersion curves), some modes appear to be influenced by weak resonances from other modes. In such cases, the axial wavenumber $k$ of the primary resonant pair for a given mode lies within the growth-rate band of another mode belonging to a different $(m_A, m_B)$ pair. We refer to these additional contributions as \textit{secondary compositions}. Table~\ref{tab:modecomp_all} lists only those secondary compositions that contribute a non-negligible amount of energy.

In some modes, the secondary contribution can be quite significant—sometimes comparable to, or even larger than, that of the primary composition. A clear example is mode A4: although its primary pair is $(-1,2)$, figure~\ref{fig:res_gw_curves_w1} shows that its axial wavenumber lies within the growth-rate band of mode A6, associated with $(0,3)$, and table~\ref{tab:modecomp_all} confirms that most of A4’s energy comes from this secondary composition.

Other modes where the secondary composition contributes significantly include A5 (A3), B4 (B6), B5 (B6), B7 (B9), and B8 (B9), with the dominant secondary mode indicated in parentheses. Please note that the axial vorticity structures that are shown in figures~\ref{fig:2dstructures1}, \ref{fig:2dstructures2}, and \ref{fig:2dstructures3}, naturally include contributions from all relevant secondary resonant pairs whenever they occur.

In table~\ref{tab:sigmafreq_all}, we list the axial wavenumber $k$, the frequencies from DNS and theory, the maximum viscous growth rates at $\Rey = 10^4$ obtained from both approaches, and the corresponding inviscid growth rates, allowing us to quantify the extent of viscous damping exhibited by each mode. Overall, the DNS results show very good agreement with the theoretical predictions. The frequencies are also in reasonably close agreement, though a consistent offset is observed: the theoretical values are systematically lower than those obtained from the DNS. At present, we do not have a clear explanation for this discrepancy.

It is also interesting to note that among the modes presented above, ring modes appear only in the B and C sets. In particular, modes B5 and B8 lie within regions where ring modes associated with $m_A$ are expected, while modes C2, C3, C4, and C5 fall within or near the corresponding ring-mode regions. Figure~\ref{fig:ringmode_structures} compares the radial distributions $|u(r)|$ versus $r$ for modes B5, B8, C2, C3, C4, and C5—extracted from DNS via Fourier decomposition of the disturbance fields—against their corresponding eigenfunctions obtained as solutions to equation~\eqref{eq22}. This comparison shows a clear agreement between the theoretical predictions and the DNS results in capturing the ring-mode nature of these modes. As seen in the plots, the eigenfunction peaks of the $m_A$ components (red curves) are shifted toward larger values of $r$, indicating that the disturbances are localized away from the vortex centre, as expected for ring modes. For a consistent comparison, the profiles are normalized by dividing all amplitudes by the maximum value of the $m_A$ component, so that the peak of $m_A$ equals unity. This same normalization factor is applied to $m_B$ for both DNS and theoretical data. Additionally, we can see from the bottom panels of (b) and (c) in figures \ref{fig:res_gw_curves_w2} and \ref{fig:res_gw_curves_w3} that the growth rate curves of these ring modes, i.e., B5, B8, C2–C5 are distinctly smaller in magnitude and narrower in width compared with those of core modes.

\subsection{Instability characteristics as a function of axial flow}
\label{subsec:theoryresults}

Having compared and validated the characteristics of the unstable modes using DNS and theory for the three discrete values of $W_0$, we now apply the theoretical framework to perform a more informative parametric study. In particular, we track each resonant pair continuously and examine how the mode characteristics and their growth rates evolve as $W_0$ varies. In this section, we carry out this analysis by following each resonance $[l_A, l_B]$ for the azimuthal wavenumber pairs $(m_A, m_B) = (-1,2)$, $(0,3)$, $(1,4)$, and $(2,5)$ over a continuous range of $W_0$ from $0$ to $-0.4$. The results are shown for the mode pairs $[1,1]$, $[2,2]$, $[1,2]$, and $[2,1]$ for all the above azimuthal combinations. For the pairs $(0,3)$ and $(1,4)$, we additionally include the results for $[1,3]$. The explicit dependence of the growth rate on $\epsilon$ and $\Rey$ in equation~\eqref{eq34Gen} also allows us to vary these parameters directly and examine how the growth rate responds. Accordingly, in the subsequent analyses, we include additional parametric studies to assess the influence of $\epsilon$ and $\Rey$ on the growth-rate behaviour.

\begin{figure}
  \centering
  \includegraphics[width=1.0\textwidth]{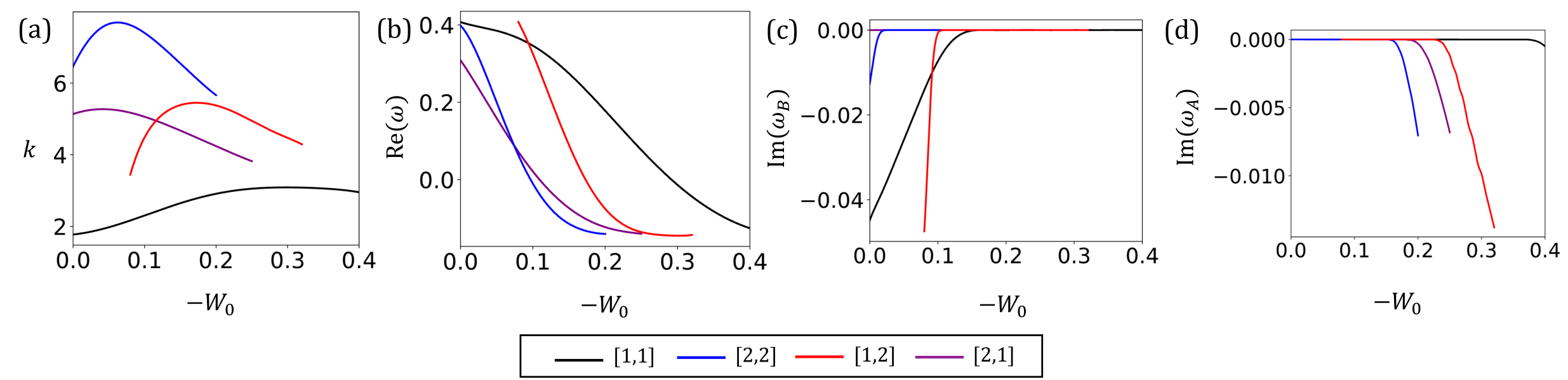}
  \caption{Characteristics of the resonant modes as a function of $W_0$ for $(m_A, m_B) = (-1, 2)$, corresponding to the branches $[l_A, l_B]$ indicated in the legend: (a) axial wavenumber $k$, (b) frequency $\Real(\omega)$, (c) damping rate of $m_B$ $\Imag(\omega_B)$, and (d) damping rate of $m_A$ $\Imag(\omega_A)$ }

\label{fig:komgsig_distr_-12}
\end{figure}

\begin{figure}
  \centering
  \includegraphics[width=1.0\textwidth]{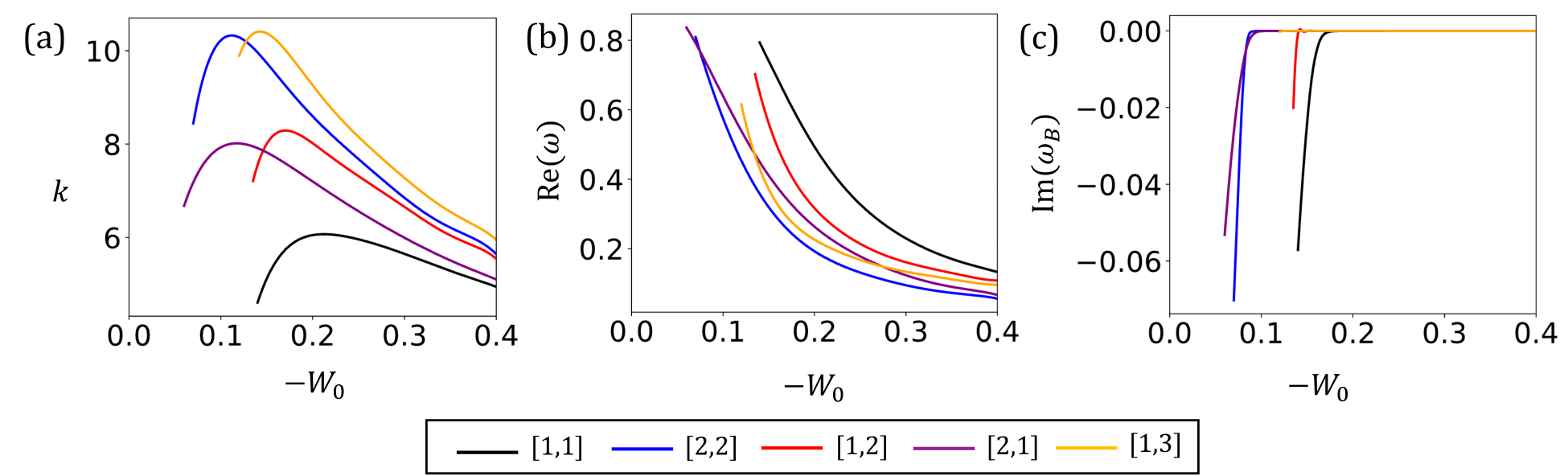}
  \caption{Same as figure \ref{fig:komgsig_distr_-12} but for $(m_A, m_B) = (0, 3)$.}
\label{fig:komgsig_distr_03}
\end{figure}

\begin{figure}
  \centering
  \includegraphics[width=1.0\textwidth]{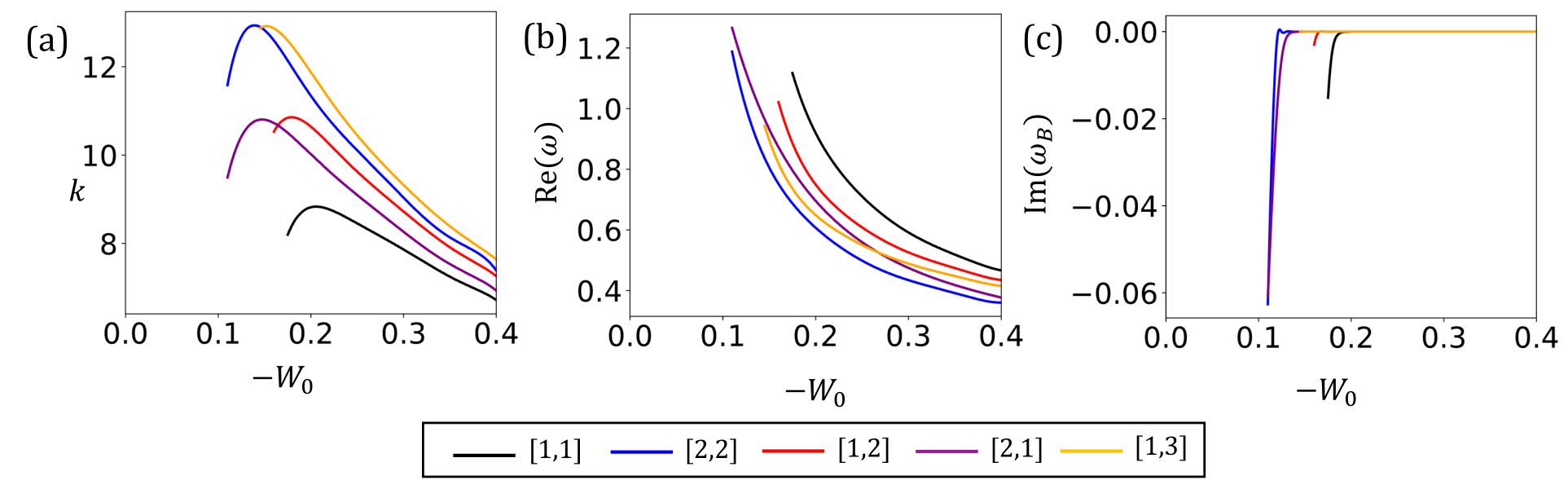}
  \caption{Same as figure \ref{fig:komgsig_distr_-12} but for $(m_A, m_B) = (1, 4)$.}
\label{fig:komgsig_distr_14}
\end{figure}

\begin{figure}
  \centering
  \includegraphics[width=1.0\textwidth]{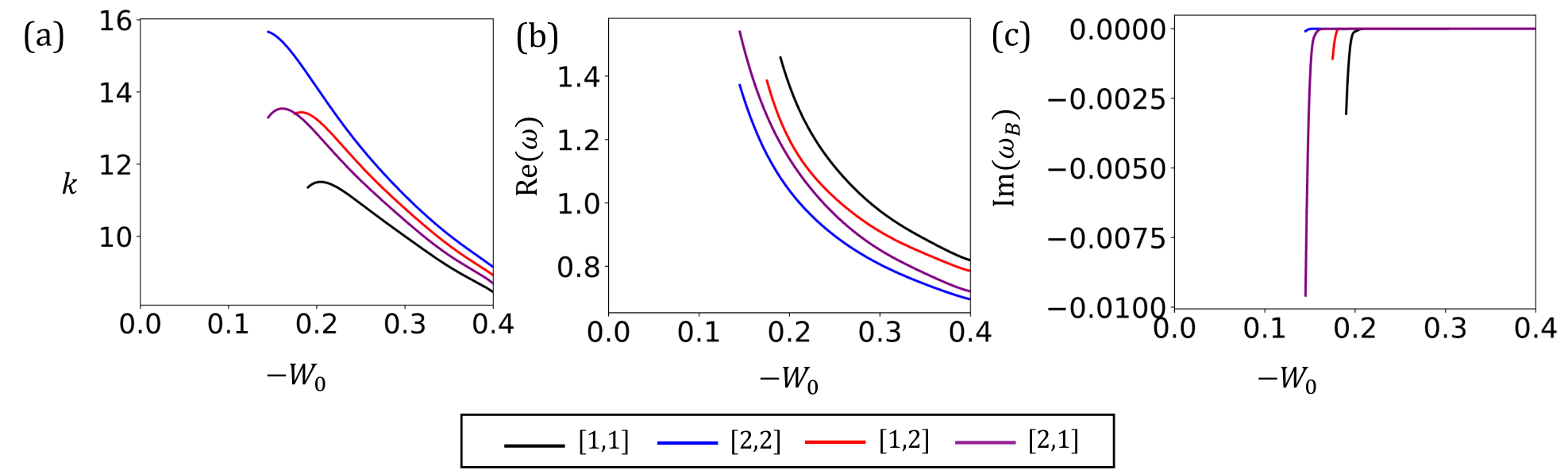}
  \caption{Same as figure \ref{fig:komgsig_distr_-12} but for $(m_A, m_B) = (2, 5)$.}
\label{fig:komgsig_distr_25}
\end{figure}

\begin{figure}
  \centering
  \includegraphics[width=1.0\textwidth]{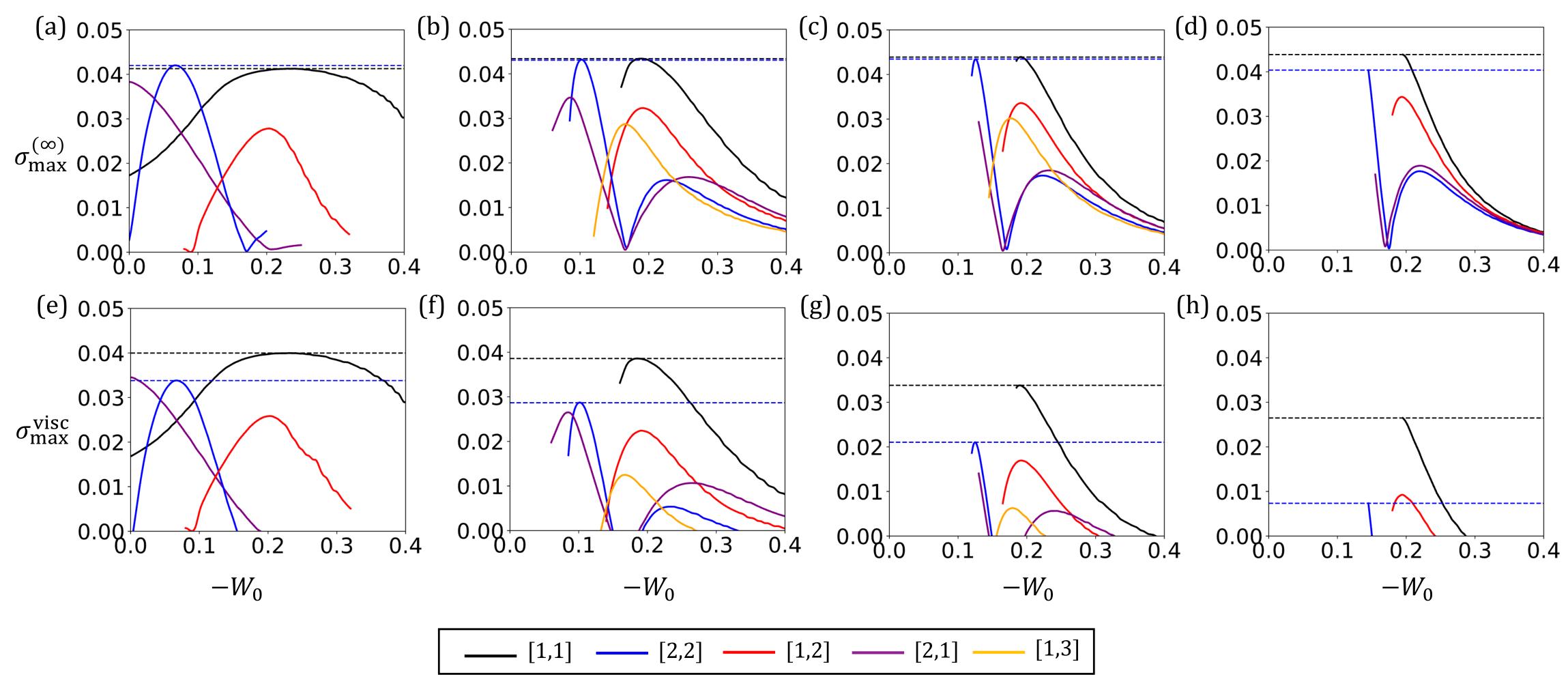}
  \caption{Panels (a), (b), (c), and (d) show the maximum inviscid growth rate as a function of $-W_0$ for (a) $(m_A, m_B) = (-1, 2)$, (b) $(0, 3)$, (c) $(1, 4)$, and (d) $(2, 5)$. Panels (e), (f), (g), and (h) display the corresponding maximum growth rates at $\Rey = 10^4$. The black dashed line indicates the maximum growth rate of the first principal mode $[1,1]$, and the blue dashed line corresponds to the second principal mode $[2,2]$.}

\label{fig:sigmaxdist}
\end{figure}

The variations of the resonant axial wavenumber $k$, resonant frequency $\Real(\omega)$, and damping rates $\Imag(\omega)$ for the $m_B$ modes are plotted as functions of $W_0$ in panels (a), (b), and (c) of figures~\ref{fig:komgsig_distr_-12}, \ref{fig:komgsig_distr_03}, \ref{fig:komgsig_distr_14}, and \ref{fig:komgsig_distr_25}. Since the $m_A$ modes become singular neutral only for the pair $(m_A, m_B) = (-1,2)$ (as discussed in section~\ref{subsec:asymptotic}), the corresponding damping rates for $m_A$ are shown exclusively for this case, in panel (d) of figure~\ref{fig:komgsig_distr_-12}. In each figure, the values of $k$ and $\Real(\omega)$ are plotted beginning from the value of $W_0$ where the corresponding resonance first appears.

It can be observed that the resonances initially exhibit significant critical-layer damping for the $m_B$ modes across all pairs. Furthermore, figure~\ref{fig:komgsig_distr_-12}(d) shows that for $(m_A, m_B) = (-1,2)$, the modes $[2,2]$, $[1,2]$, and $[2,1]$ terminate as the critical-layer damping of $m_A$ becomes significant at higher values of $|W_0|$. It is also noteworthy that the mode $(-1,2,[1,1])$ exhibits the most extended region of critical-layer damping across the range of $W_0$, from $W_0 = 0$ to approximately $W_0 = -0.15$, as seen in figure~\ref{fig:komgsig_distr_-12}(c). Additionally, the figures show that the resonant frequency decreases monotonically as $|W_0|$ increases, appearing to asymptotically approach lower values for all modes. This trend indicates that the modes oscillate more slowly as the axial flow strengthens. It is also noteworthy that for the pair $(m_A, m_B) = (-1,2)$, the frequency changes sign beyond a certain value of $W_0$, implying that only the $(-1,2)$ modes can become stationary for some axial flow strength.

Next, we computed the growth rates of all the above modes, using equation~\eqref{eq34Gen}. The computation uses the eigenvalues $(k, \Real(\omega), \Imag(\omega))$ and the corresponding eigenfunctions and adjoint modes at resonance, as described in section \ref{subsec:theorymath}. The integrals appearing in the inner products of equations~\eqref{eq30}, \eqref{eq31}, and \eqref{eq32} are evaluated along a complex contour that avoids the critical point associated with the $m_B$ mode for all azimuthal wavenumber pairs considered. In the case of $(-1,2)$, the contour also avoids the critical point of the $m_A$ mode, since for certain values of $W_0$, even $m = -1$ exhibits a critical point on the real axis, as discussed in section~\ref{subsec:asymptotic}. Modes that do not have a critical point are also integrated along the same contour, as the results remain unchanged.

\begin{figure}
  \centering
  \includegraphics[width=1.0\textwidth]{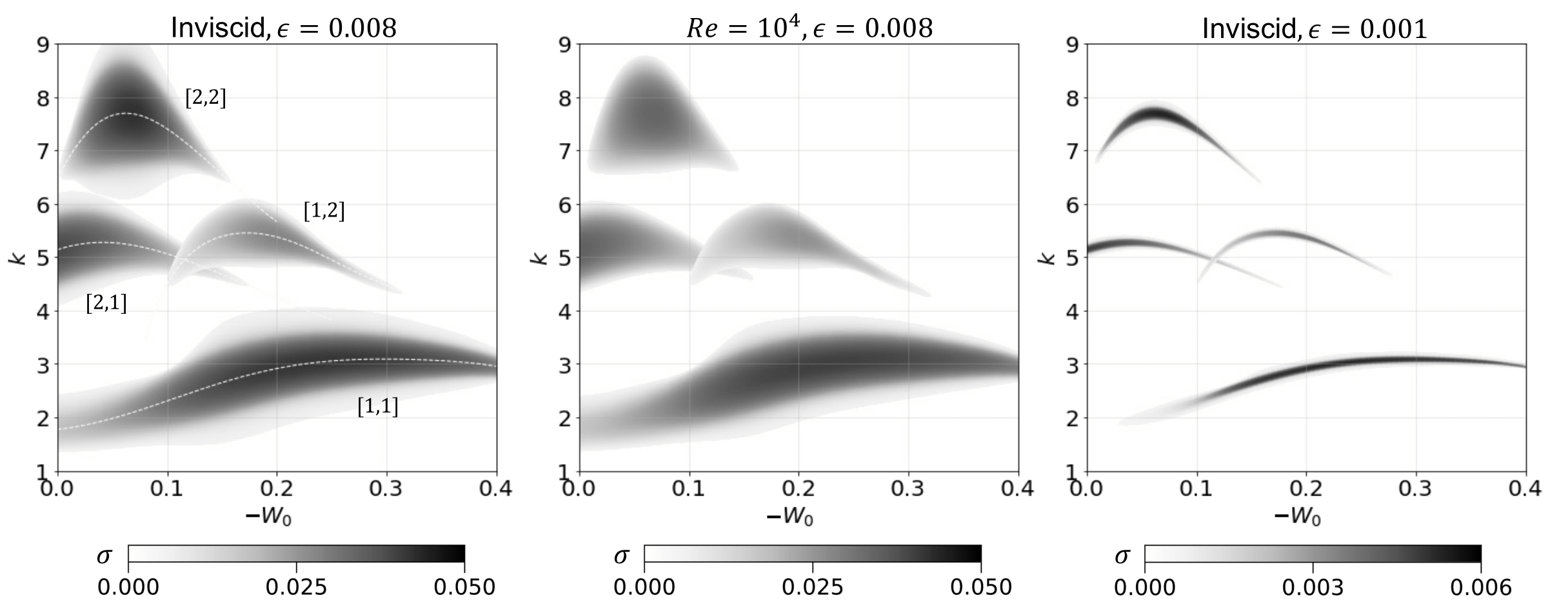}
  \caption{Growth rate $\sigma$ contours on the $(k, W_0)$ plane for $(m_A, m_B) = (-1, 2)$, corresponding to mode branches $[l_A, l_B]$ as labeled near each island. Left panel: inviscid case with $\epsilon = 0.008$; centre panel: viscous case with $\Rey = 10^4$ and $\epsilon = 0.008$; right panel: inviscid case with $\epsilon = 0.001$. The corresponding resonant axial wavenumbers as a function of $-W_0$ for those modes are also plotted as thin white dashed lines for reference.}

\label{fig:gwcontouts_-12}
\end{figure}

\begin{figure}
  \centering
  \includegraphics[width=1.0\textwidth]{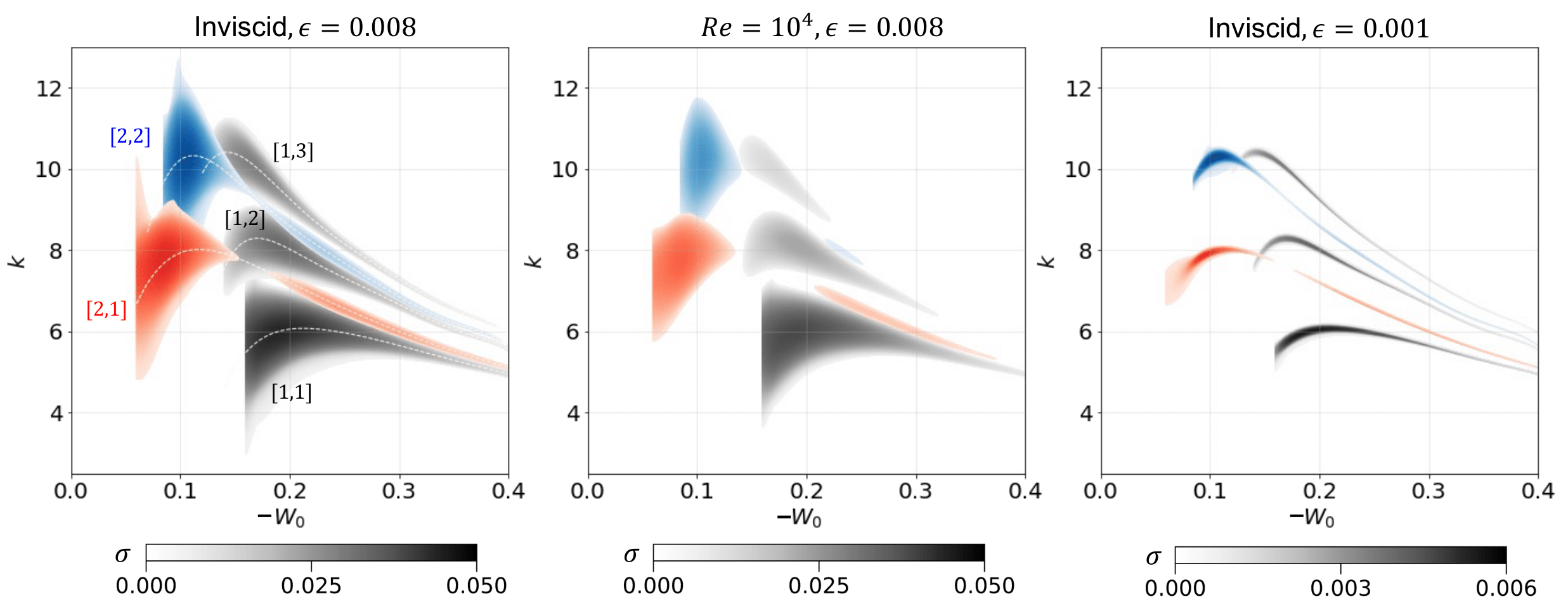}
  \caption{Growth rate $\sigma$ contours on the $(k, W_0)$ plane for $(m_A, m_B) = (0, 3)$, corresponding to mode branches $[l_A, l_B]$ as labeled near each island. Left panel: inviscid case with $\epsilon = 0.008$; centre panel: viscous case with $\Rey = 10^4$ and $\epsilon = 0.008$; right panel: inviscid case with $\epsilon = 0.001$. The modes $[2,1]$ and $[2,2]$, which exhibit two local maxima in growth rate and distinct 'tail' regions, are highlighted in red and blue, respectively, for clarity. The corresponding resonant axial wavenumbers as a function of $-W_0$ for those modes are also plotted as thin white dashed lines for reference.}
\label{fig:gwcontouts_03}
\end{figure}

\begin{figure}
  \centering
  \includegraphics[width=1.0\textwidth]{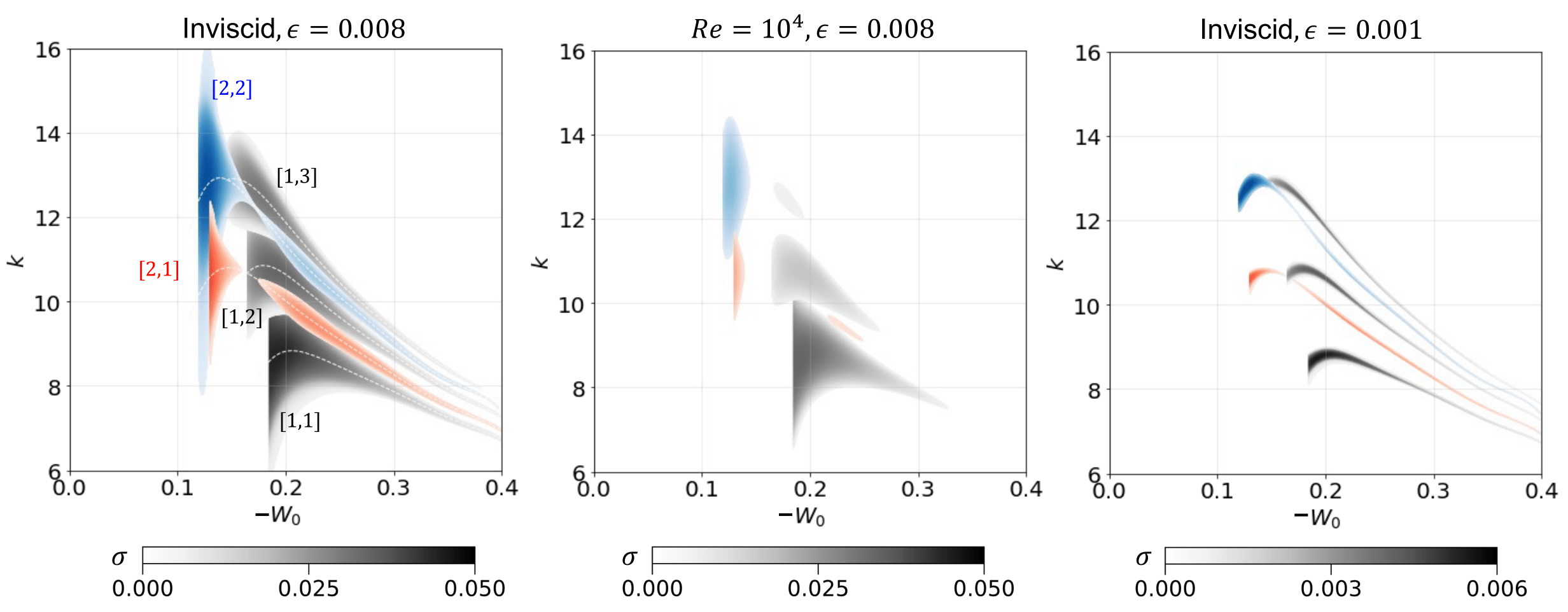}
  \caption{Growth rate $\sigma$ contours on the $(k, W_0)$ plane for $(m_A, m_B) = (1, 4)$, corresponding to mode branches $[l_A, l_B]$ as labeled near each island. Left panel: inviscid case with $\epsilon = 0.008$; centre panel: viscous case with $\Rey = 10^4$ and $\epsilon = 0.008$; right panel: inviscid case with $\epsilon = 0.001$. The modes $[2,1]$ and $[2,2]$, which exhibit two local maxima in growth rate and distinct 'tail' regions, are highlighted in red and blue, respectively, for clarity. The corresponding resonant axial wavenumbers as a function of $-W_0$ for those modes are also plotted as thin white dashed lines for reference.}
\label{fig:gwcontouts_14}
\end{figure}

\begin{figure}
  \centering
  \includegraphics[width=1.0\textwidth]{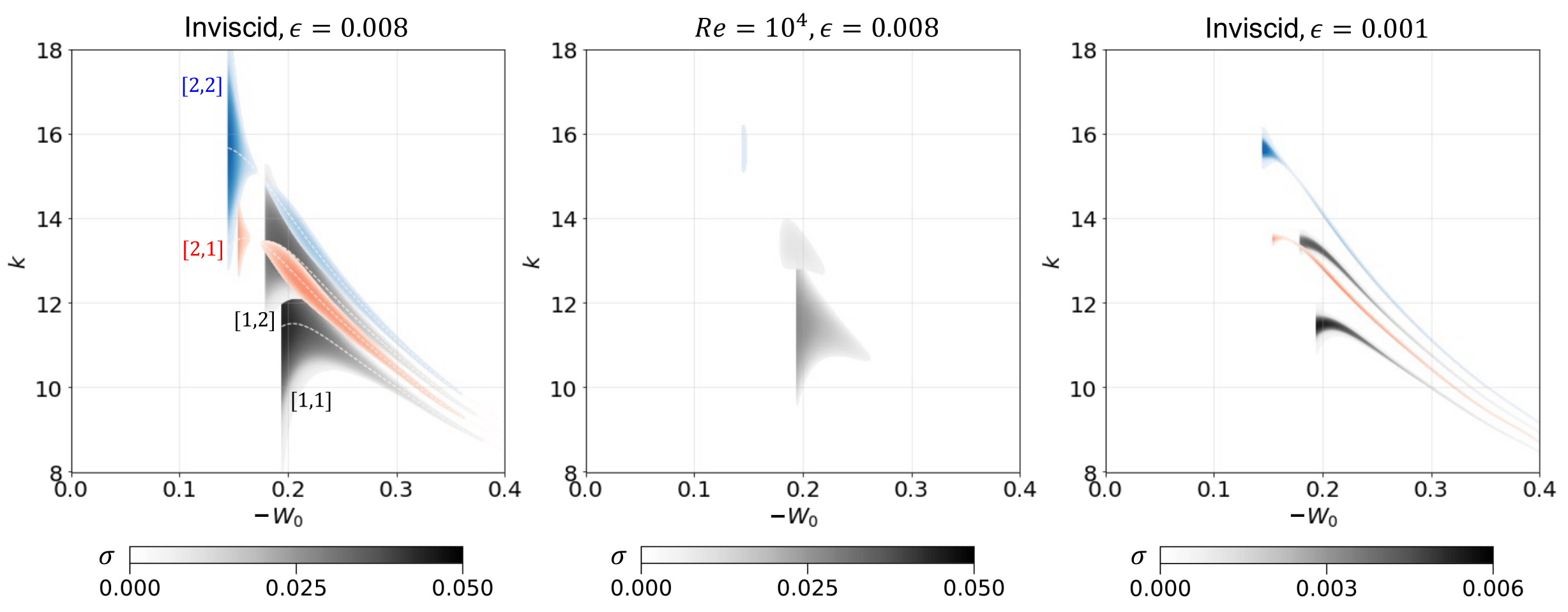}
 \caption{Growth rate $\sigma$ contours on the $(k, W_0)$ plane for $(m_A, m_B) = (2, 5)$, corresponding to mode branches $[l_A, l_B]$ as labeled near each island. Left panel: inviscid case with $\epsilon = 0.008$; centre panel: viscous case with $\Rey = 10^4$ and $\epsilon = 0.008$; right panel: inviscid case with $\epsilon = 0.001$. The modes $[2,1]$ and $[2,2]$, which exhibit two local maxima in growth rate and distinct 'tail' regions, are highlighted in red and blue, respectively, for clarity. The corresponding resonant axial wavenumbers as a function of $-W_0$ for those modes are also plotted as thin white dashed lines for reference.}
\label{fig:gwcontouts_25}
\end{figure}

\begin{figure}
  \centering
  \includegraphics[width=1.0\textwidth]{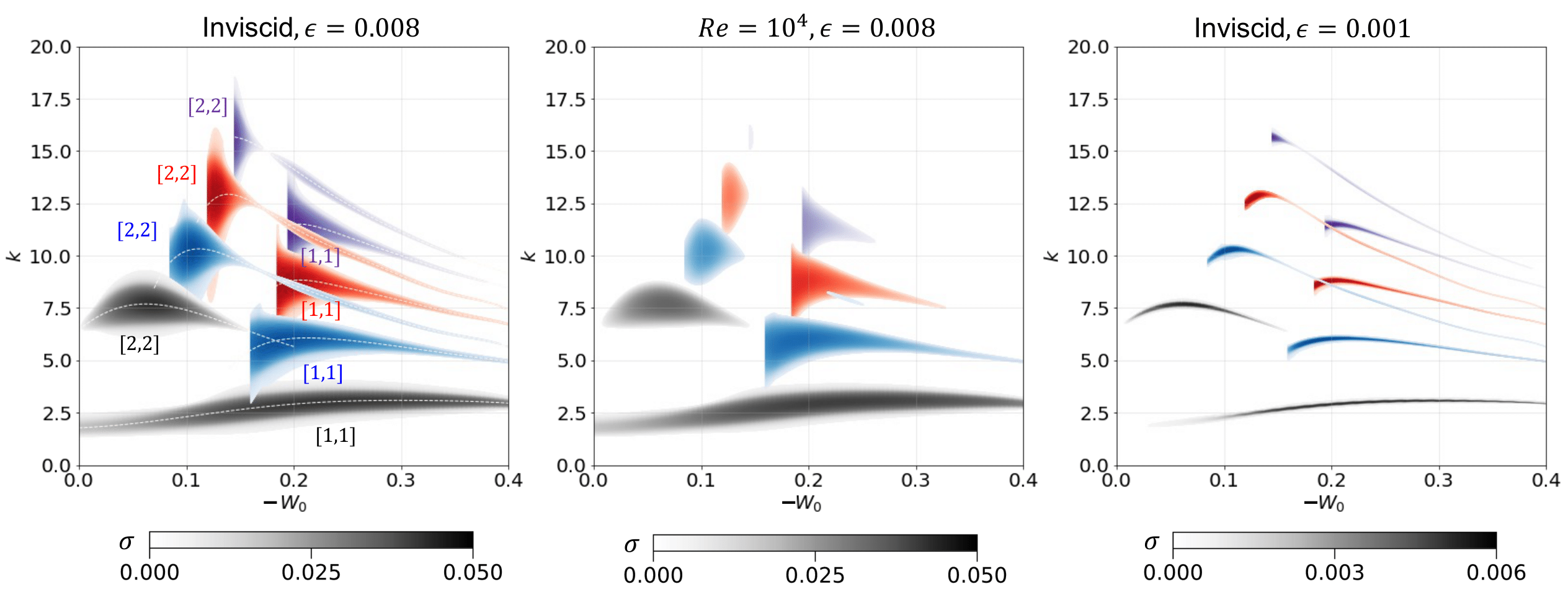}
 \caption{Growth rate $\sigma$ contours on the $(k, W_0)$ plane for the principal modes $[1,1]$ and $[2,2]$, shown for $(m_A, m_B) = (-1, 2)$ (grey), $(0, 3)$ (blue), $(1, 4)$ (red), and $(2, 5)$ (purple). Left panel: inviscid case with $\epsilon = 0.008$; centre panel: viscous case with $\Rey = 10^4$ and $\epsilon = 0.008$; right panel: inviscid case with $\epsilon = 0.001$.}

\label{fig:gwcontouts_princip}
\end{figure}

Figure~\ref{fig:sigmaxdist} shows the maximum inviscid growth rates in panels (a)-(d) and viscous growth rates at $\Rey = 10^4$ in (e)-(h) for all modes as functions of $W_0$. The progressive decrease in viscous growth rates from $(-1,2)$ to $(2,5)$ clearly illustrates the expected increase in volumic viscous damping with higher values of $m$ and $k$. Figures~\ref{fig:gwcontouts_-12}, \ref{fig:gwcontouts_03}, \ref{fig:gwcontouts_14}, and \ref{fig:gwcontouts_25} show contour plots of the growth-rate magnitudes across the $k$-bands centered around their respective resonant axial wavenumbers $k_c$, plotted as functions of $W_0$. In each figure, the left and middle panels compare the contours for the inviscid case and for $Re = 10^4$ at $\epsilon = 0.008$. The right panels display the corresponding contours at $\epsilon = 0.001$, highlighting how the growth rates change under weaker straining. As observed, the growth rates become substantially smaller (scaling with $\epsilon^2$), and the associated $k$-bands narrow considerably when $\epsilon = 0.001$. The growth rate $\sigma$ for all modes generally reaches a peak at a specific value of $W_0$ and then decreases as $W_0$ increases further.

An important observation that can be made by looking at the growth rate contours is that for several resonant modes—such as the $[1,1]$ and $[2,1]$ modes of $(0,3)$ in figure~\ref{fig:gwcontouts_03}, for example, the contours begin with a sharp vertical separation at a specific $W_0$, where the growth-rate bands are relatively wide. To the left of this $W_0$ value, the effective critical-layer damping, computed using equations~\eqref{eq36a} and \eqref{eq36b}, becomes positive near the resonant $k_c^{(\infty)}$, producing unphysical growth-rate bands that are discarded. This behaviour is expected because, near this cut-off and to its left, the two Kelvin modes—when viewed on the dispersion curves—are nearly tangent and only barely form a crossing point. As a result, the expression for growth rate in equation~\eqref{eq34Gen} is not expected to give accurate values in this region unless higher-order terms are included.

Overall, the principal modes $[1,1]$ and $[2,2]$ exhibit higher peak growth rates than the other modes across all cases in the inviscid limit. This trend is consistent with previous observations for both the Rankine vortex and the elliptic instability. For illustrative purpose, in figure~\ref{fig:gwcontouts_princip}, we plot the growth rate contours for only the principal modes across all azimuthal wavenumber pairs analyzed.

\begin{figure}
  \centering
  \includegraphics[width=1.0\textwidth]{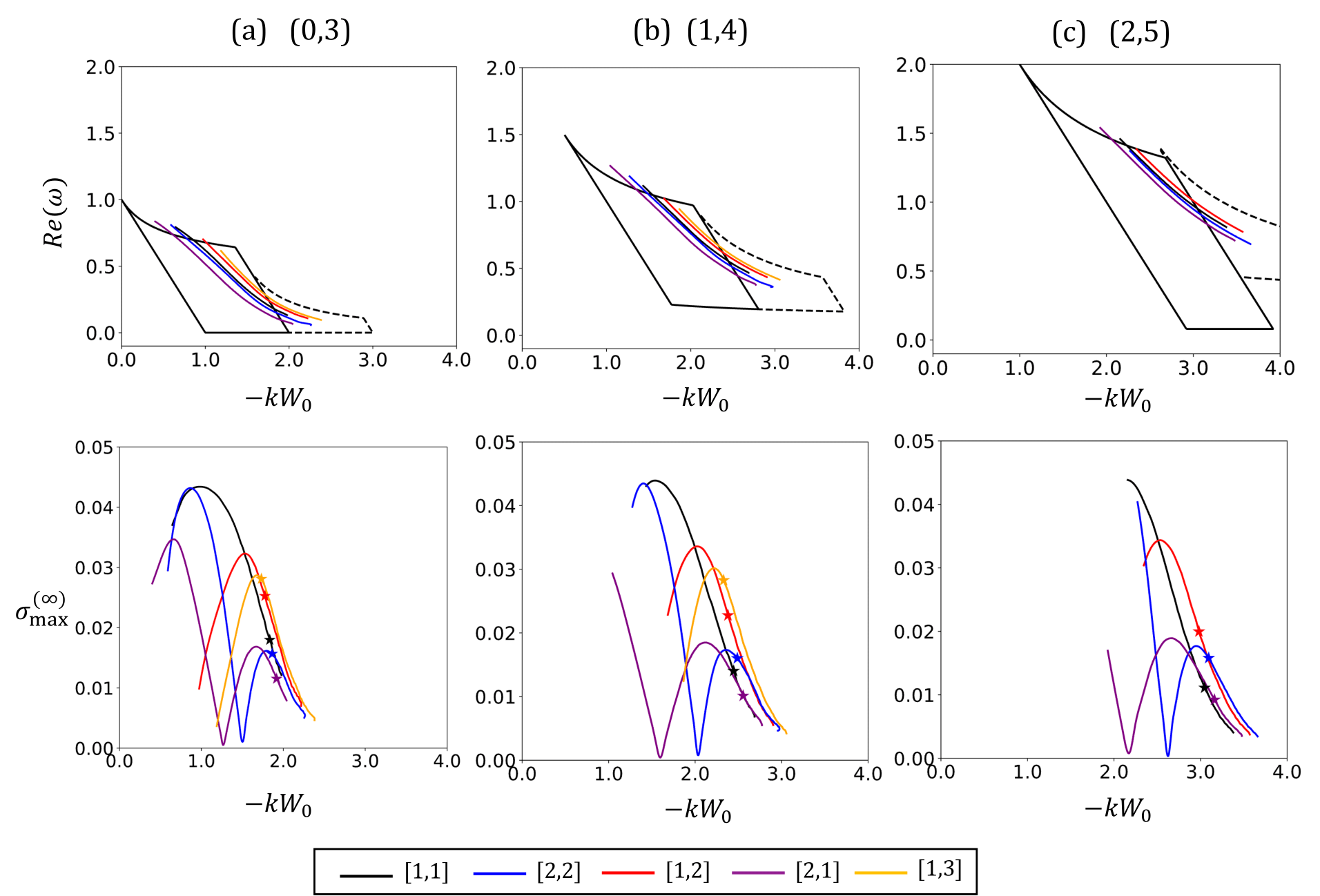}
  \caption{Transition of the modes corresponding to $m_A$ from core modes to ring modes for (a) $(m_A, m_B) = (0, 3)$, (b) $(1, 4)$, and (c) $(2, 5)$. Top panels: frequency versus $-kW_0$ for the resonant modes, overlaid on the core and ring mode domains predicted by asymptotic theory. Bottom panels: maximum inviscid growth rate versus $-kW_0$, with stars indicating the transition points from core to ring modes. Labels and color scheme follow those used in figures~\ref{fig:komgsig_distr_-12}–\ref{fig:sigmaxdist}.}

\label{fig:ringmodes_thcurves}
\end{figure}

\begin{figure}
  \centering
  \includegraphics[width=1.00\textwidth]{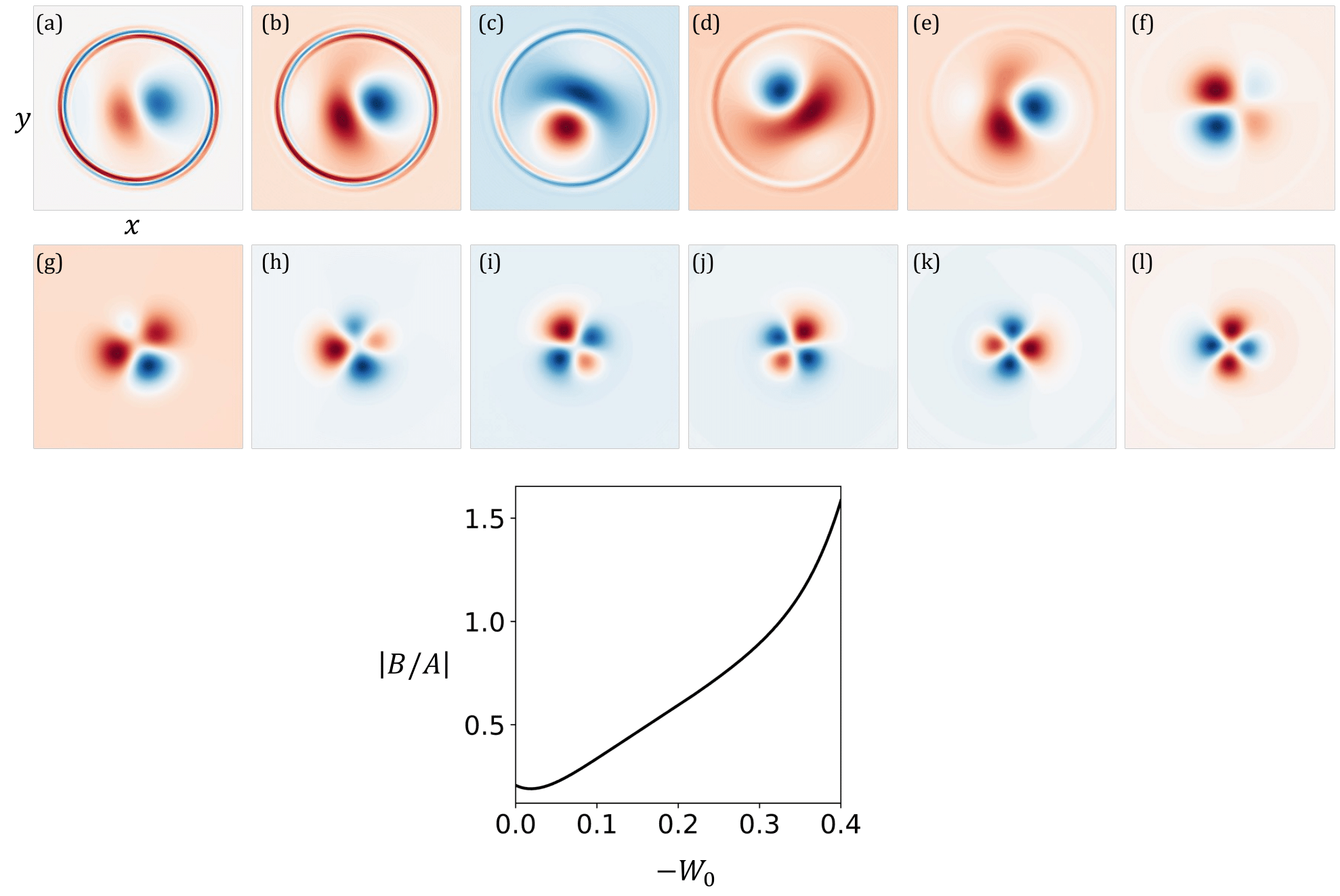}
  \caption{Top: Evolution of the 2-D axial disturbance vorticity field of the mode $(-1,2,[1,1])$, obtained theoretically at $\Rey = 10^4$, as $W_0$ varies from $0$ to $-0.4$. Panels (a) to (l) correspond to values of $-W_0$ equal to $0$, $0.025$, $0.05$, $0.075$, $0.1$, $0.15$, $0.2$, $0.25$, $0.3$, $0.35$, $0.375$, and $0.4$, respectively. Bottom: Amplitude ratio of $m_B$ to $m_A$ as a function $W_0$.}

\label{fig:evolstructures_m-12}
\end{figure}

\begin{figure}
  \centering
   \includegraphics[width=1.00\textwidth]{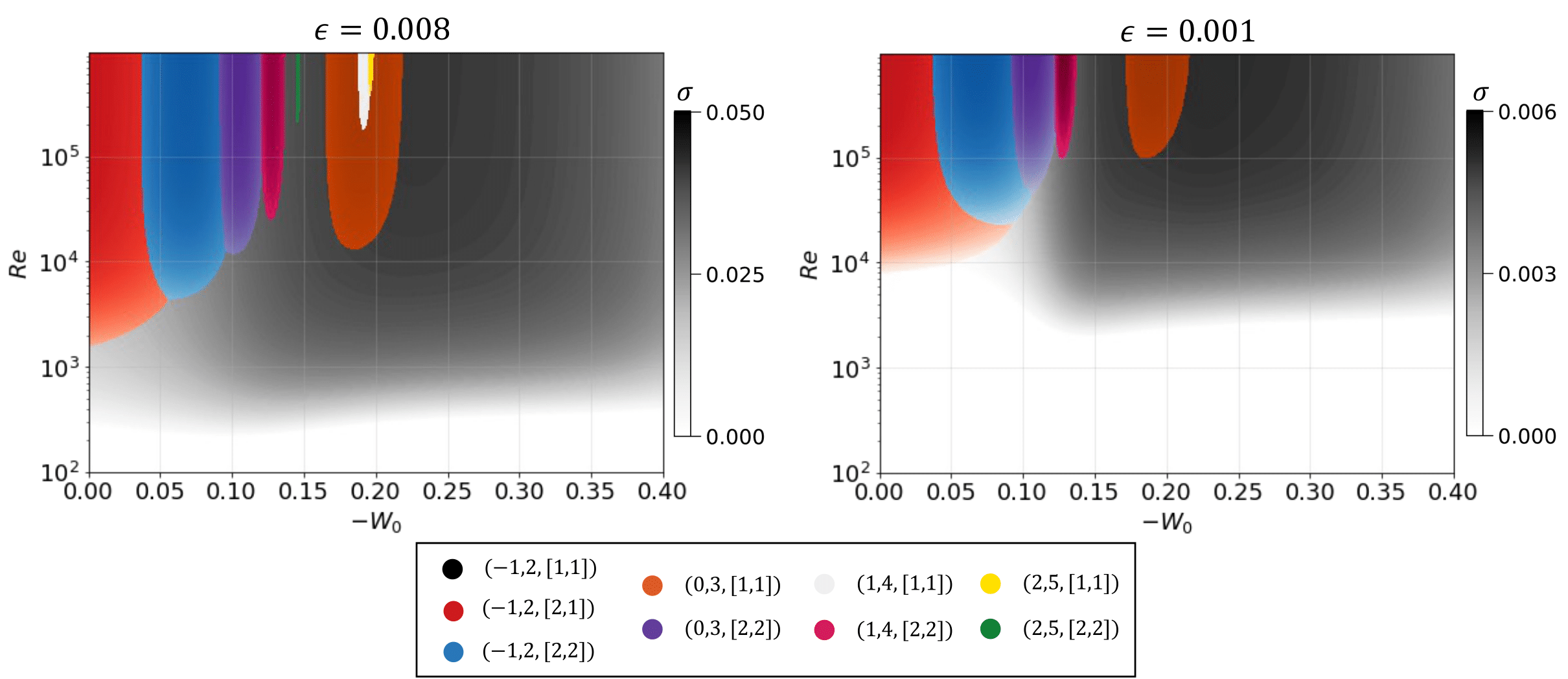}
  \caption{Contours of the maximum growth rate in the $(W_0, \Rey)$ parameter space, computed for different modes identified by $(m_A, m_B, [l_A, l_B])$ as indicated in the legend. The left panel corresponds to $\epsilon = 0.008$, and the right panel to $\epsilon = 0.001$.}

\label{fig:gwcontouts_Re}
\end{figure}

\begin{figure}
  \centering
  \includegraphics[width=0.9\textwidth]{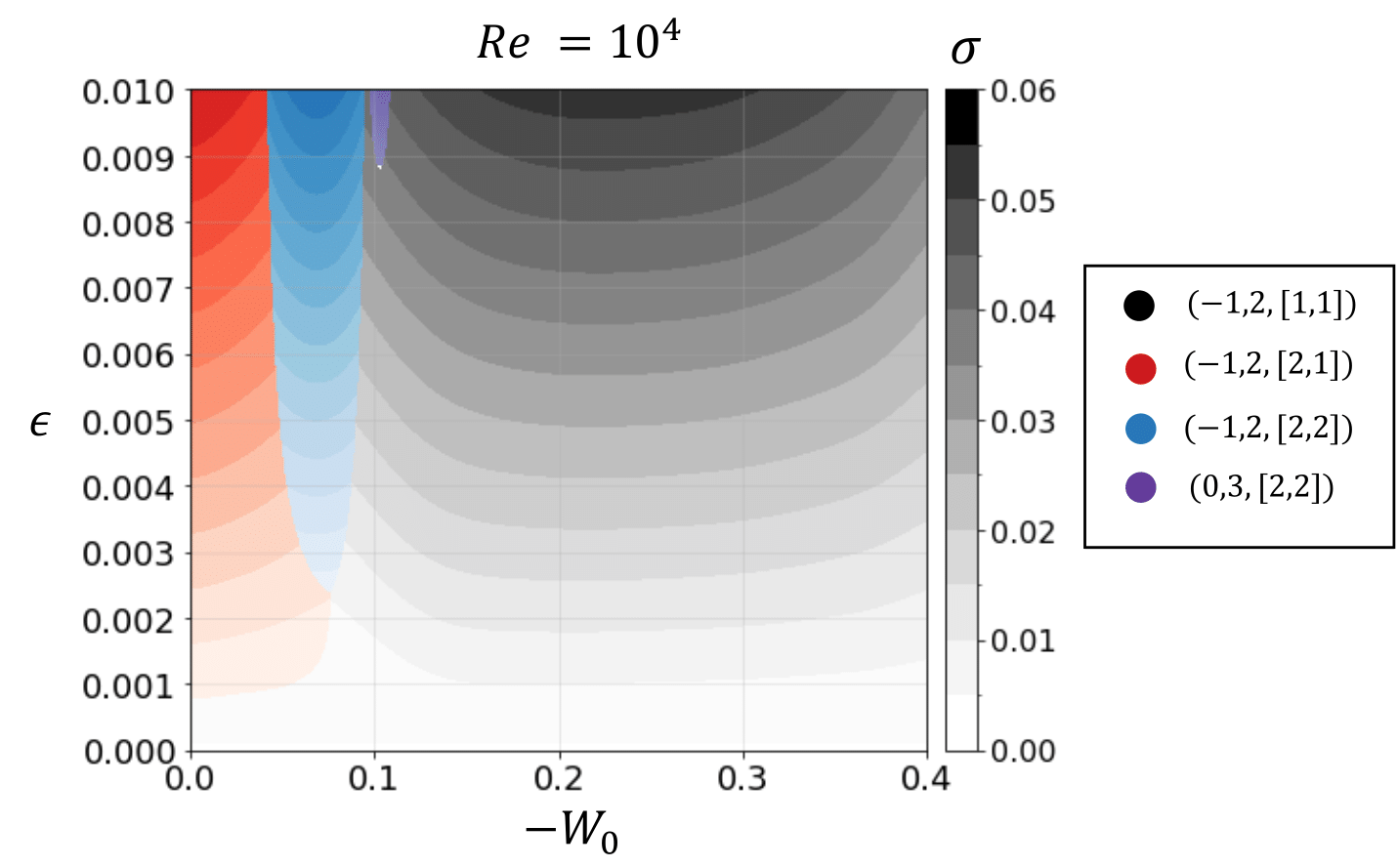}
  \caption{Contours of the maximum growth rate in the $(W_0, \epsilon)$ parameter space for $\Rey = 10^4$. Modes identified by $(m_A, m_B, [l_A, l_B])$ are indicated in the legend.}

\label{fig:gwcontouts_Ep}
\end{figure}

We shall now discuss several other interesting observations and inferences that can be drawn from these results.
\begin{enumerate}

\item The modes corresponding to $m_A$ in the resonant pairs $(0,3)$, $(1,4)$, and $(2,5)$ transition into ring modes at certain values of $kW_0$, as discussed in section~\ref{subsec:asymptotic}. This transition for the modes in the present analysis is illustrated more clearly in figure~\ref{fig:ringmodes_thcurves}. Interestingly, across all three pairs, the order in which the ring-mode transition occurs among the branch combinations follows a consistent pattern: $[1,3]$, $[1,2]$, $[1,1]$, $[2,2]$, $[2,1]$. Modes involving the first branch of $m_A$ are the first to undergo the transition, and the transition order is also correlated with higher branch numbers of $m_B$. As seen in the growth rate contours in figures~\ref{fig:gwcontouts_03}, \ref{fig:gwcontouts_14}, and \ref{fig:gwcontouts_25}, these ring-mode branches are characterized by narrow, elongated $k$-bands and relatively low growth rates.

\item Another interesting observation is found for the modes $[2,1]$ and $[2,2]$, where the resonance involves the second branch of $m_A$. For these modes, the growth rate first reaches a local maximum as $W_0$ increases, then drops to nearly zero, and subsequently rises again to a second local maximum before eventually vanishing. This two-peak structure is not observed in modes from other branches. At first glance, one might suspect that the second peak is related to the onset of ring-mode behaviour; however, this is unlikely, as the bottom panels of figure~\ref{fig:ringmodes_thcurves} clearly show that the transition from core to ring modes occurs well after the second rise in growth. At present, we therefore do not have a definitive theoretical explanation for this peculiar feature.

\item We can verify from figures~\ref{fig:sigmaxdist}(a),(e) and~\ref{fig:gwcontouts_-12} that the Lamb–Oseen vortex results reported in AHL25 are fully recovered at $W_0 = 0$. At $\Rey = 10^4$, only the modes $[1,1]$ and $[2,1]$ corresponding to $(m_A, m_B) = (-1, 2)$ exhibit positive growth rates, whereas the $[2,2]$ mode has only a very small growth rate in the inviscid limit. As expected, at $W_0 = 0$ the growth rate of the $[1,1]$ mode is lower than that of $[2,1]$, owing to the strong critical-layer damping affecting the $m = 2$ component. However, as $|W_0|$ increases, the growth rate of $(-1,2,[1,1])$ rises substantially, reaching its maximum for $|W_0| \approx 0.2$–$0.3$. This trend can be attributed to the reduction in critical-layer damping of the $m = -2$ mode, which decreases steadily and approaches zero as the axial flow magnitude increases (see figure~\ref{fig:komgsig_distr_-12}(c)). 
This point is further illustrated by examining the evolution of the 2-D axial disturbance vorticity field of the $(-1,2,[1,1])$ mode. In figure~\ref{fig:evolstructures_m-12}, the top panel shows this evolution as $W_0$ varies from $0$ to $-0.4$, computed from the eigenfunctions of equation~\eqref{eq22} at $\Rey = 10^4$ and evaluated using equation~\eqref{eq24} with $t = 0$ and $z = 0$. At small $|W_0|$, a clear spiral pattern—stemming from the critical-layer wave associated with $m = 2$—is visible. As $|W_0|$ increases and the critical-layer influence weakens, the spiral gradually disappears, and the mode transitions to a more regular structure.
The bottom panel of figure~\ref{fig:evolstructures_m-12} shows the associated variation of the amplitude ratio $|B/A|$ between the $m_B$ and $m_A$ components, computed using equations~\eqref{eq28} or~\eqref{eq29}. The ratio increases monotonically with $|W_0|$, indicating growing participation of the $m_B$ mode as its critical-layer damping decreases. Once $|W_0|$ exceeds a moderate threshold, the ratio surpasses unity, indicating that $m_B$ overtakes $m_A$ and becomes the dominant contributor to the mode composition.

The variation of the maximum growth rate magnitudes on $\Rey$ is analyzed in the $\Rey$--$W_0$ parameter space and shown in figure~\ref{fig:gwcontouts_Re}, for $\epsilon = 0.008$ and $0.001$. It is evident that the mode $(-1,2,[1,1])$, corresponding to the first principal mode of $(m_A, m_B) = (-1, 2)$, dominates most of the $\Rey$--$W_0$ parameter space. In contrast, other principal modes only become the most unstable in limited regions, particularly at high Reynolds numbers and small values of $-W_0$, less than approximately $0.2$. Beyond $W_0 = -0.2$, the mode $(-1,2,[1,1])$ becomes the most unstable mode again until it fades away for some value beyond $W_0 = -0.4$. Therefore, the mode $(-1,2,[1,1])$ clearly emerges as the most unstable, especially at larger values of $W_0$. To ensure the reliability of this comparison across Reynolds numbers, we restrict the upper limit of $\Rey$ to $10^6$. The reason is that, in principle, as one approaches the inviscid limit, modes associated with azimuthal pairs higher than $(m_A, m_B) = (2,5)$ may also become unstable, though at much larger axial wavenumbers. By limiting the range to $\Rey \le 10^6$, such higher-order modes are expected to remain strongly suppressed by volumic viscous damping and therefore do not influence the trends shown in the figure. 

\item In figure~\ref{fig:gwcontouts_Ep}, we examine the variation of the maximum growth rates across the $\epsilon$--$W_0$ parameter space at a fixed Reynolds number of $\Rey = 10^4$. Three modes—$(-1,2,[1,1])$, $(-1,2,[2,1])$, and $(-1,2,[2,2])$—can be seen to dominate over a broad range of $\epsilon$. At $W_0 = 0$, the most unstable mode is $(-1,2,[2,1])$, which then briefly gives way to $(-1,2,[2,2])$, before ultimately transitioning to $(-1,2,[1,1])$ as the axial flow magnitude increases. Overall, these results indicate that triangular instability at moderate Reynolds numbers (e.g.\ $\Rey \approx 10^4$ as in laboratory conditions) can persist even under very weak straining, with instability observable down to $\epsilon \approx 0.001$ and growth rates on the order of $\sigma \approx 5\times 10^{-3}$, before becoming negligible at still weaker straining strengths.

\end{enumerate}

\section{Conclusions}
\label{sec:summary}

We carried out a theoretical and numerical study of the triangular instability of a Batchelor vortex subjected to a stationary triangular strain field generated by three symmetrically positioned satellite vortices. The main aim was to examine how axial flow influences the characteristics of this instability. Theoretically, the triangular instability results from a resonant interaction between two quasi-neutral Kelvin modes of the unstrained vortex, with azimuthal wavenumbers $m$ and $m+3$, coupled to the background triangular strain field. A multiscale analysis of the linearized perturbation equations was used to derive the corresponding growth rate. In the numerical simulations, we identified the most unstable mode by solving the linearized Navier–Stokes equations around a fixed base flow.

In our earlier study \citep{triangjfm2025}, we demonstrated the existence of triangular instability and characterized it for the Lamb–Oseen vortex, where resonance was limited to the azimuthal wavenumber pair $(m_A, m_B) = (-1, 2)$ (or its symmetric counterpart), with all other pairs suppressed by strong critical-layer damping. In the present work, we showed that introducing axial flow alters the resonance behaviour, and for $W_0$ in the range $[0, -0.4]$, additional pairs such as $(0, 3)$, $(1, 4)$, and $(2, 5)$ also become resonant. We presented the results in two parts: (i) For $W_0 = -0.1$, $-0.2$, and $-0.3$, we computed the growth rates of unstable modes numerically and compared them with theoretical predictions for the corresponding resonant modes. The simulations were conducted at $\Rey = 10^4$ and $\epsilon = 0.008$, showing excellent agreement between theory and numerical simulations. We also provided the axial disturbance vorticity structures and the energy ratios of the participating modes. (ii) For a few of the resonant modes of the azimuthal wavenumber pairs $(-1, 2)$, $(0, 3)$, $(1, 4)$, and $(2, 5)$, we tracked the respective mode characteristics, including growth rates, across a continuous range of $W_0$ from $0$ to $-0.4$. The key findings are summarized in the next paragraph.

In \citet{triangjfm2025}, the most unstable mode for the Lamb--Oseen vortex in an inviscid setting or moderately high Reynolds numbers was found to be $(-1,2,[2,1])$, formed by the second branch of $m=-1$ and the first branch of $m=2$, and shown to dominate most of the $(\Rey,\epsilon)$ parameter space. The mode $(-1,2,[1,1])$ was reported to become dominant only at low Reynolds numbers. In the present work, however, we have shown that as the axial flow becomes stronger, the resonance associated with the branch pair $[2,1]$ weakens, and the dominance shifts—first to $[2,2]$ and ultimately to $[1,1]$—as a consequence of the reduction in critical-layer damping. Although principal modes from other pairs such as $(0, 3)$, $(1, 4)$, and $(2, 5)$ become most unstable over limited ranges for values of $W_0$ less than about $-0.2$, the mode $(-1, 2, [1, 1])$ eventually overtakes all others and remains dominant across a broader parameter space, thus emerging as the most dominant mode. We also showed that the modes associated with $m_A$ in the pairs $(0, 3)$, $(1, 4)$, and $(2, 5)$ undergo a transition to ring modes beyond certain values of $W_0$. These ring modes are marked by narrow growth-rate bands in $k$ and smaller magnitudes of growth rates.

Another useful result is that, using the asymptotic framework described in section~\ref{subsec:asymptotic}, one can find that no possible resonant mode overlaps exist for differences in azimuthal wavenumbers greater than three. This can also be verified by observing the absence of crossing points in the dispersion curves of their Kelvin modes computed numerically. This suggests that instabilities beyond triangular instability are unlikely. While detailed numerical and experimental studies would be needed to definitively rule out higher-order instabilities due to parametric resonance, it appears reasonable to conclude that in an arbitrary 2-D Lamb–Oseen or Batchelor vortex equilibrium, elliptic instability typically arises at second-order strain, and triangular instability at third-order. Therefore, elliptic instability modes are generally expected to dominate, with triangular instability modes contributing to a lesser extent. However, triangular instability can become the primary mechanism when the flow is near a third-order symmetrical state, or in a perfectly third-order symmetrical configuration like the one studied here. Additionally, since our investigation is restricted to linear stability, these conclusions hold only in the absence of nonlinear effects; incorporating nonlinear dynamics could significantly alter this picture. 

In this study, although small values of axial flow were considered to ensure that viscous centre modes do not become more unstable than the triangular instability modes, it is important to verify whether any viscous centre modes become unstable within the range of $W_0$ explored here, and how their characteristics compare to those of triangular instability—particularly at higher magnitudes of $W_0$. Using the asymptotic theory developed by \citet{le2007large}, it is found that in the limit of infinite Reynolds number, viscous centre modes with azimuthal wavenumber $m \geq 1$ are always unstable for $W_0 < 0$. These modes have axial wavenumbers approximately equal to $-mW_0$, and their growth rates scale as $Re^{-1/3}$, making them negligible at very high Reynolds numbers. However, at lower Reynolds numbers, their growth rates increase, and the critical magnitude of $W_0$ above which they become unstable also increases. As an example, we examine the case of $\Rey = 10^4$, and compute the critical axial flow values beyond which viscous centre modes with $m > 0$ become unstable, using the estimates provided in \citet{fabre2008viscous}. For $m = 1$, the critical value is $W_0 \approx -0.33$; for $m = 2$, $W_0 \approx -0.406$; and for $m = 3$, $W_0 \approx -0.541$. Therefore, within the range $-0.4 \leq W_0 \leq 0$ considered in this study, only the $m = 1$ viscous centre mode becomes unstable, and only when $W_0 < -0.33$. The numerically determined magnitude of the critical value for $m = 1$ by \citet{fabre_viscous_2004} is slightly lower, at $W_0 \approx -0.31$. At $W_0 = -0.4$ and $Re = 10^4$, the numerical growth rate of the $m = 1$ viscous centre mode is $0.004$, whereas the most unstable triangular instability mode $(-1,2,[1,1])$ at the same conditions and $\epsilon = 0.008$ has a growth rate of $0.0292$, which is approximately $7.3$ times larger. The critical value of $\epsilon$ at which the centre mode overtakes the triangular instability mode in growth rate is approximately $\epsilon \approx 0.00135$.
As for the inviscid centre modes, the condition given by \citet{heaton2007centre} indicates that they become weakly unstable only when $|W_0| > 0.433$, which lies outside the range of $W_0$ considered in this study.

Overall, the results of this work can have useful implications in studies concerning rotor wakes of propellers with three blades, where the hub vortex can in principle be subject to a third-order strain induced by the satellite vortices, and opening the door for potential triangular instability modes to develop in the wake of the hub vortex. Also, triangular instability can be one of the potential mechanisms, along with elliptic instability, which can be used to explain the pathways of transition to turbulence. However, several questions need to be addressed that merit further exploration in future work. 

In \citet{triangjfm2025} and this study, we analyzed the triangular instability characteristics under a stationary strain field, where the circulations of the satellite vortices were set to cancel out their mutual rotation. However, in a general case, the circulations may differ, causing the satellite vortices to rotate around the hub vortex. Previous work by \citet{dizes_non-axisymmetric_2000} showed that such rotation introduces an additional critical layer in the base flow of the strained vortex. The influence of rotation on the characteristics of the elliptic instability has also been investigated in \citet{sipp1999vortices, le2000three, le2002theoretical}.

The base flow in our model is simplified to two dimensions. However, in real rotor wakes, the tip vortices form helical structures, which generate a third-order azimuthal strain field with helical symmetry and axial variation linked to the helical pitch. Investigating how this axial dependence influences triangular instability would be a valuable direction for future work. Additionally, the hub vortex is not always aligned with the rotor axis—it can adopt a helical geometry depending on the rotor design \citep{DuranVenegas19}. Such configurations are susceptible to additional short-wavelength instabilities, such as elliptic and curvature instabilities \citep{blanco-rodriguez_elliptic_2016, blanco-rodriguez_curvature_2017}. Triangular instability may also compete with instabilities developing in the cores of the helical satellite vortices and with global long-wavelength instabilities that affect the full hub–satellite system \citep{Quaranta15, DuranVenegas19}. Recently, \citet{hattori_effect_2024} explored the effect of pitch on the nonlinear dynamics of the long-wave instability of helical vortices. Other important studies addressing rotor wake systems comprising the hub vortex, root vortices, and tip vortices include \citet{felli2011mechanisms, sorensen2011instability, iungo2013linear, viola2014prediction, ashton2016hub, felli2018propeller, posa2022dynamics}, among others. These works have shown evidence of \citet{widnall_stability_1972} type instabilities in rotor wakes and determined that the stability of the hub vortex is influenced by the evolving stability of the tip vortices. Single-helix ($m = 1$) and double-helix ($m = 2$) modes have also been identified as dominant modes of the hub vortex, arising from unstable inviscid perturbations in a rotating vortex with axial flow, where stability is given by the Leibovich–Stewartson criterion. We suggest that the findings of our study, along with future extensions on triangular instability, can contribute to refining the understanding of the complex interplay of instabilities and the rich dynamics occurring in rotor wakes, with potential benefits for the design of wind farms, ship propellers, and other rotating machinery.
\\\\
\textbf{Declaration of interests.} The authors report no conflict of interest.
\\
\appendix

\section{Linear stability analysis -- operators and coefficients}\label{appA}
The operators appearing in equation~\eqref{eq20} are
 \begin{equation}
\mathcal{L} = \begin{pmatrix}
1 & 0 & 0 & 0\\ 
0 & 1 & 0 & 0\\ 
0 & 0 & 1 & 0\\ 
0 & 0 & 0 & 0
\end{pmatrix},
\label{eqa1}
\end{equation}
\begin{equation}
\mathcal{P} = \begin{pmatrix}
W_0e^{-r^2} & 0 & 0 & 0\\ 
0 & W_0e^{-r^2} & 0 & 0\\ 
0 & 0 & W_0e^{-r^2} & 1\\ 
0 & 0 & 1 & 0
\end{pmatrix},
\label{eqa2}
\end{equation}
\begin{equation}
\mathcal{M} = \begin{pmatrix}
\Omega\frac{\partial }{\partial \theta} & -2\Omega & 0 & \frac{\partial }{\partial r} \\
 2\Omega + r\frac{d\Omega}{dr}& \Omega\frac{\partial }{\partial \theta} & 0 & \frac{1}{r}\frac{\partial }{\partial \theta} \\
 -2r W_0e^{-r^2} & 0 & \Omega\frac{\partial }{\partial \theta} & 0 \\
 \frac{\partial }{\partial r} + \frac{1}{r} & \frac{1}{r}\frac{\partial }{\partial \theta} & 0 & 0
\end{pmatrix},
\label{eqa3}
\end{equation}
\begin{equation}
\mathcal{V} = \begin{pmatrix}
\Delta - \frac{1}{r^2} & -\frac{2}{r^2}\frac{\partial }{\partial \theta} & 0 & 0 \\
\frac{2}{r^2}\frac{\partial }{\partial \theta} & \Delta - \frac{1}{r^2} & 0 & 0 \\
0 & 0 & \Delta & 0 \\
0 & 0 & 0 & 0
\end{pmatrix},
\label{eqa4}
\end{equation}
where
\begin{equation}
\Delta = \frac{1}{r}\frac{\partial }{\partial r} + \frac{\partial^2 }{\partial r^2} + \frac{1}{r^2}\frac{\partial^2 }{\partial \theta^2} + \frac{\partial^2 }{\partial z^2}.
\label{eqa5}
\end{equation}
\begin{equation}
\mathcal{N} = \frac{1}{2}\begin{pmatrix}
N_{11} & N_{12} & 0 & 0 \\
 N_{21}& N_{22} & 0 & 0 \\
 N_{31} & N_{31} & N_{33} &  0\\
0 & 0 & 0 & 0
\end{pmatrix},
\label{eqa6}
\end{equation}
with
\begin{align}
N_{11} &= \frac{3f}{r^2} - \frac{3f'}{r} - \frac{3f}{r}\frac{\partial }{\partial r} - \frac{\mathrm{i}f'}{r}\frac{\partial }{\partial \theta} + \frac{2W_0kr^2f}{1-e^{r^2}}, \label{eqa7} \\
N_{12} &= -\frac{9\mathrm{i}f}{r^2} + \frac{2\mathrm{i}f'}{r}, \label{eqa8} \\
N_{21} &= -\frac{\mathrm{i}f'}{r} - \mathrm{i}f'', \label{eqa9}\\
N_{22} &= -\frac{3f}{r^2} + \frac{3f'}{r} - \frac{3f}{r}\frac{\partial }{\partial r} - \frac{\mathrm{i}f'}{r}\frac{\partial }{\partial \theta} + \frac{2W_0kr^2f}{1-e^{r^2}}, \label{eqa10} \\
N_{31} &= -2\mathrm{i}W_0\left( \frac{2rf + 2e^{r^2}r^3f + r^2f'}{1-e^{r^2}}\right), \label{eqa11} \\
N_{32} &=  \frac{6W_0rf}{1-e^{r^2}}, \label{eqa12} \\
N_{33} &= - \frac{3f}{r}\frac{\partial }{\partial r} - \frac{\mathrm{i}f'}{r}\frac{\partial }{\partial \theta} + \frac{2W_0kr^2f}{1-e^{r^2}}. \label{eqa13}
\end{align}

Here $\overline{\mathcal{N}}$ is the complex conjugate of $\mathcal{N}$.

\bibliographystyle{jfm}
\bibliography{jfm.bib}

\end{document}